\documentclass[ preprint,12pt]{elsarticle}
\usepackage{array}
\usepackage{amsmath,amssymb}
\usepackage{breqn}
\usepackage{subcaption}
\usepackage{subcaption}
\usepackage{stackengine}
\usepackage{verbatim} 
\usepackage{algorithm}
\usepackage{algorithmic}
\usepackage[usenames,dvipsnames]{color} 
\newcolumntype{R}{>{\centering\arraybackslash}m{8cm}}
\usepackage{float}
\usepackage{lineno,hyperref}
\modulolinenumbers[5]
\journal{arXiv}

\begin{document}

\begin{frontmatter}

\title{Optimal Workplace Occupancy Strategies during the COVID-19 Pandemic}

\author[a,b]{Mansoor Davoodi\corref{cor1}}
\ead{m.davoodi-monfared@hzdr.de}

\author[a,b]{Abhishek Senapati}
\ead{a.senapati@hzdr.de}

\author[a,b]{Adam Mertel}
\ead{a.mertel@hzdr.de}

\author[a,b]{Weronika Schlechte-Welnicz}
\ead{w.schlechte-welnicz@hzdr.de}

\author[a,b,c,d]{Justin M. Calabrese}
\ead{j.calabrese@hzdr.de}


\cortext[cor1]{Corresponding author.}
\address[a]{Center for Advanced Systems Understanding (CASUS), G$\ddot{o}$rlitz, Germany}
\address[b]{Helmholtz-Zentrum Dresden-Rossendorf (HZDR), Dresden, Germany}
\address[c]{Helmholtz Centre for Environmental Research-UFZ, Leipzig, Germany}
\address[d]{Dept. of Biology, University of Maryland, College Park, MD, USA}

\begin{abstract}
During the COVID-19 pandemic, many organizations (e.g. businesses, companies, government facilities, etc.) have attempted to reduce infection risk by employing partial home office strategies. However, working from home can also reduce productivity for certain types of work and particular employees.
Over the long term, many organizations therefore face a need to balance infection risk against productivity. Motivated by this trade-off, we model this situation as a bi-objective optimization problem and propose a practical approach to find trade-off (Pareto optimal) solutions. We present a new probabilistic framework to compute the expected number of infected employees as a function of key parameters including: the incidence level in the neighborhood of the organization, the COVID-19 transmission rate, the number of employees, the percentage of vaccinated employees, the testing frequency, and the contact rate among employees. 
We implement the model and the optimization algorithm and perform several numerical experiments with different parameter settings. Furthermore, we provide an online application based on the models and algorithms developed in this paper, which can be used to identify the optimal workplace occupancy rate for real-world organizations.
\end{abstract}


\end{frontmatter}


\section{Introduction}

The COVID-19 pandemic has caused worldwide social and economic disruptions over the past two years. One consequence has been that companies, organizations, and government facilities (hereafter ``organizations'') have suffered reductions in the quality and quantity of the services they provide \cite{yang2021effects, avdiu2020face, dingel2020many,arntz2020working,bloom2020impact}. This decreased efficiency is a direct consequence of the need to manage infection risk in the workplace. As COVID-19 spreads mostly through direct/close contacts between individuals, the risk of infection tends to increase with workplace occupancy rate. To manage infection risk, many organizations have implemented a range of strategies including: (\textit{i}) adopting recommended personal hygiene measures (e.g., wearing a mask, social distancing, etc.), (\textit{ii}) reducing the number of employees present in the workplace via teleworking, (\textit{iii}) screening employees via periodic testing, and (\textit{iv}) encouraging employees to get vaccinated \cite{world2020coronavirus,ho2020workplace, cirrincione2020covid}. Though these strategies have worked quite well to manage workplace infection risk, they also come with certain costs and limitations. For example, telework has the potential to completely eliminate workplace infection risk, but it may also decrease productivity for particular tasks or individuals, and may not possible at all for certain types of work~\cite{yang2021effects, tavares2020teleworking}. It is therefore evident that a trade-off between productivity and infection risk exists, and an important question that arises here is \textit{“what is the optimal presence rate in the workplace during the pandemic?”}.


Motivated by these circumstances, we develop a model of disease transmission within organizations that allows us to balance the trade-off between productivity and infection risk. Our model incorporates the local incidence level and inherent transmission rate of COVID-19 as well as key within-organization factors such as the testing frequency, rate of vaccination, and contact rate among the employees. Additionally, our model accounts for differences in efficiency between working remotely and being present in the workplace. This factor varies from organization to organization depending on the type of work they do and their inputs, processes and outputs. Hereafter, we refer to the ratio of efficiency at home to efficiency at office as the \textit{productivity} for simplicity.

We employ a probabilistic framework that accounts for the aforementioned influential factors and allows us to compute the expected number of infected employees at any time after an infection is introduced into the facility. This framework has the advantages of being simple, fast, and practical, while still being customizable for specific organizations. 
Moreover, our model can be efficiently used for any organization size. Based on this model, we also develop an optimization application to find Pareto optimal solutions of the twin objectives of maximizing productivity and minimizing infection risk in organizations. The application allows the user to adjust all input parameters and immediately visualize the results of any change.

This paper is organized in five sections. In the second section, we review related studies. In the third section, we introduce the core model parameters and formally state the problem. In addition, we propose the probabilistic model and an algorithm to compute infection risk and the Pareto optimal solutions. In the fourth section, we perform several numerical experiments and test the ability of the model and algorithm to identify Pareto optimal solutions. In the fifth and final section, we draw conclusions and outline future directions.

\section{Related Work}
Managing infection risk within organizations has been recognized as a key theme in responding to and living with the pandemic. Consequently, many mathematical models have been developed specifically to investigate several aspects of transmission mechanisms and control strategies for COVID-19 in organizations \cite{wells2021optimal,ingram2021covid,evans2021impact,biala2022efficient,furati2021fractional}. In addition, several quantitative studies have also been carried out to gain insight on the outcomes of strategies proposed for different organization types\cite{marshall2020sentinel,herstein2021characteristics,gunawardana2021longitudinal,erber2020strategies}. Most of these studies focused on implementing strategies for organizations like hospitals, health care facilities, nursing homes, offices, and schools. For example, \cite{chin2020frequency, see2021modeling} have focused on identifying the optimal frequency of testing in healthcare environments and nursing homes, while others have demonstrated that immediate testing of symptomatic individuals and quarantining of their primary contacts via contact tracing can potentially reduce disease transmission among health workers and other high risk groups \cite{grassly2020comparison}. Still other surveillance-focused studies have explored the performance of non-adaptive combinatorial group testing \cite{mcdermott2021nonadaptive}, and the importance of self-isolation and contact tracing measures \cite{kucharski2020effectiveness}. Finally, \cite{hernandez2020optimal} proposed an optimal testing strategy to minimize the presence of pre-symptomatic and asymptomatic employees in the workplace. 

Some studies have investigated how scheduling personnel in the workplace can minimize the consequences of infection.
In this context, a \textit{desynchronization strategy} has been proposed in which the workers in the healthcare system are divided into two non-overlapping teams who will be working in alternating weeks \cite{sanchez2021modelling}. Similarly, \cite{sanchez2021regular} studied the effectiveness of regular testing and a \textit{desynchronization} protocol in preventing COVID-19 infection in hospitals while accounting for both internal and external sources of infection. The effect of a cyclic 4-day work and 10-day lockdown strategy in the workplace has been investigated by~\cite{karin2020adaptive}. In nursing homes and long-term care facilities, \cite{lucia2021modeling} constructed a mathematical model based on bipartite networks consisting of health care workers and facility residents to investigate how restructuring interactions via the concept of  \textit{shied immunity} can change outbreak dynamics. Specifically, shield immunity refers to the practice of proportionally increasing the interactions between recovered and susceptible individuals to protect the susceptible from infection. Additionally, \cite{iavicoli2021risk} developed a decision support system for different employment sectors like agriculture, manufacturing, construction, and human health in Italy and measured the occupational risk of infection in workplace for each sector based on type of work-activity, involvement of third parties in the work processes, and risk of social aggregation.


Several recommendations based on mathematical models have been proposed in reopening different activities in workplaces. For instance, some studies have explored how enhanced levels of testing, contact-tracing, adoption of facial masks, and home quarantine can facilitate relaxation of social distancing interventions~\cite{aleta2020modelling,d2020restart}. 
Similarly, \cite{meier2020working} studied how designing an optimal employee screening strategy can reduce workplace infections, as well as how such a strategy affects the feasibility of return to work policies. 

From the above discussion, it is clear that most of these studies have focused on modeling the impact of implementing different infection reduction measures in the workplace. Additionally, some studies have recommended strategies for personnel scheduling and reopening different activities. However, no study to date has modeled the basic mechanics of workplace disease transmission while accounting for testing frequency, the contact rate, the vaccination rate, and the differing efficiencies of home vs office work. 



\section{Modeling and Solution Approach for the Workplace Presence Problem}


While teleworking has advantages and disadvantages \cite {tavares2017telework, nakrovsiene2019working}, the COVID-19 pandemic has forced many organizations to massively increase the amount of remote work they allow. The realized productivity of employees working from home depends strongly on the type of tasks they perform \cite {choudhury2021work}. However, for many job types (e.g., hospitality, banking, information technology) productivity increases with the physical presence of employees in the workplace \cite {battiston2017distance}. So, while working from home clearly reduces infection risk, we conclude that it will also decrease productivity for many types of work \cite {farooq2021potential, beno2021data}. Therefore, a trade-off, mediated by workplace presence rate, will often exist between productivity and infection risk.  

In this section, we formulate this trade-off as a bi-objective optimization problem to find the optimal rate of workplace presence while also considering the risk of infection at the facility. Notably, if the employee productivity ratio of some organization is not decreased by telework (e.g., due to a 100\% digital workflow), then clearly no trade-off between workplace infection risk and productivity exists and the optimal (trivial) strategy is minimum workplace occupancy. While some chance of infection for employees working from home still exists, organizations do not have direct control over this, and there is no disease spread in the workplace. Thus, in this paper, we only consider organizations for which total productivity increases with increasing workplace occupancy.


Table \ref{table_params} defines the model's key parameters, variables, and assumptions.

\begin{table}[h!]
  \begin{center}
    \caption{The parameters and variables used in the proposed model.}
    \label{table_params}
    \begin{tabular}{l|l} \hline
      \textbf{Notation} & \textbf{Definition} \\\hline
      $n$ & The total number of employees \\
      $n_v$ & The number of vaccinated employees \\
	$n_I$ & The number of infected employees \\
	$\beta_u$ & The COVID-19 transmission rate for unvaccinated individuals\\
	$\beta_v$ & The COVID-19 transmission rate for vaccinated individuals\\
	$prod$  & (home) Productivity ratio \\
	$\tau$  & The average interval between tests of the same employee \\
	$\rho$  & Probability of an infection arriving at the facility \\
	$\kappa$  & The average number of contacts per employee per day \\
	$occup$  & The workplace occupancy rate \\
    \end{tabular}
  \end{center}
\end{table}

\textbf{Parameters, Variables, and Assumptions of Model}
\begin{itemize}
\item $n$: The total number of employees who regularly work in the facility 
\item $n_v$: The number of employees who are (fully) vaccinated with one of the available COVID-19 vaccines. We count partially
vaccinated individuals (e.g., those who have only had one dose) as non-vaccinated. 
\item $n_I$: The expected number of infected employees that can spread the disease in the facility. As soon as they test positive or the disease's symptoms appear, they are quarantined and no longer counted as an active employee.
\item $\beta_u$ and $\beta_v$: The per contact probability of virus transmission from one infected individual to one unvaccinated ($\beta_u$) or one vaccinated ($\beta_v$) individual.
\item $prod$: The productivity ratio is the productivity of working at home compared to working at the office. This factor varies from facility to facility and from employee to employee, based on their mission and the type of services they provide. Managers can determine this factor based on their experience in working under the COVID-19 pandemic and comparing it with normal situations. Precisely,
$prod= \frac{Productivity~of~working~from~Home}{Productivity~of~working~at~Office}$.
So, $prod=1$ means that there is no difference between productivity for working at home and workplace, and $prod=0$ means the employees has no productivity when they work from home.
\item $\tau$: The time interval between two sequential tests of the same employee, averaged over all employees. As the symptoms of COVID-19 appear within two weeks \cite{paul2020distribution, qin2020estimation}, we assume $\tau$ varies from one day to 14 days (\textit{incubation period}). In order to keep the model as simple and parameter sparse as possible, we assume perfect tests with no false positives or negatives. This would not qualitatively change the results, and we could easily incorporate the testing error as a coefficient in computing the expected time to detect the infection.
\item $\rho$: The per day probability that an infection arrives at the facility. This can be computed by considering the number of employees in the facility and the incidence level in the neighborhood of the facility. Specifically, we estimate this probability based on the reported number of infections over the last 7 days in the focal region (e.g., see https://www.coronavirus.sachsen.de).

\item $\kappa$: The average number of contacts per employee per day. We assume the employees and clients mostly wear masks and keep a distance of at least 1.5 meters.
\item $occup$: The rate (percentage) of employees present in the workplace. This is the only decision variable of the model, and should be determined optimally.
\end{itemize}

As previously mentioned, all variables and parameters are given except the decision variable $occup$, which is the output of the model. Furthermore, we divide the employees into three groups: ($i$) those who work at the facility, ($ii$) those who work from home, and ($iii$) those who are infected. The productivity ratio of the first group is one, and those of second and third groups are $prod$ and zero, respectively. So, $n-n_I$ is the number of healthy and active employees and of them $occup$ percent are at the workplace with full productivity, and the remaining $1-occup$ percent work from home with productivity $prod$. Therefore, the total productivity of an organization can be determined by the following formula\footnote {It is possible to customize this formula and define it based on the outcomes in a company if a clear definition is available. In this paper, we skip such details and only focus on this basic and general definition of the total productivity.}
\begin{equation}
Total~Productivity=occup\times(n - \mathbb{E}(n_I)) + prod\times(1-occup)\times(n-\mathbb{E}(n_I))
\label{prod_obj}
\end{equation}

Therefore, one objective function of the model is \textbf{maximizing the total productivity}, and the other one is \textbf {minimizing the expected number of infections}, $\mathbb{E}(n_I)$. There is a clear conflict between these two objectives. Note that, $prod$ is the productivity ratio determined by the decision-maker. Without loss of generality, we consider $prod \in [0,1)$. Note that, for $prod \ge 1$, there is a trivial optimal solution $occup=0$, that is working from home is strictly preferred to working at the office, and it results in minimizing the number of infected employees. The most important part of the objective functions is computing the expected number of infected employees, $\mathbb{E}(n_I)$, after the arrival of an infection. Indeed, we must compute it as a function of time and of the influential parameters mentioned in Table \ref{table_params}, which are explained in detail below.

\subsection{Probabilistic framework for computing the number of infected employees}
We start with a full contact network of unvaccinated employees to explain our approach. We then extend it to accommodate both vaccinated and unvaccinated employees. All possible pairs of employees are equally likely to contact each other, but the average number of contacts per employee per day is bounded by $\kappa$.  Suppose an infection arrives at the facility at time (day) $t=0$. We then compute the probability of an arbitrary individual being infected after $\Delta t$ days, $P_I(\Delta t, n, \beta_u,\kappa)$, when the disease transmission rate is $\beta_u$ and the number of contacts per day is $\kappa$. This probability function allows one to easily compute the expected number of infected individuals as

\begin{equation}
\mathbb{E} \left[ (n_I(\Delta t,n,\beta_u,\kappa) \right]=1+ (n-1)\times P_I(\Delta t,n, \beta_u,\kappa).
\label{Exp_n_I}
\end{equation}

\begin{figure}[t]
\centering
\includegraphics[width=4cm]{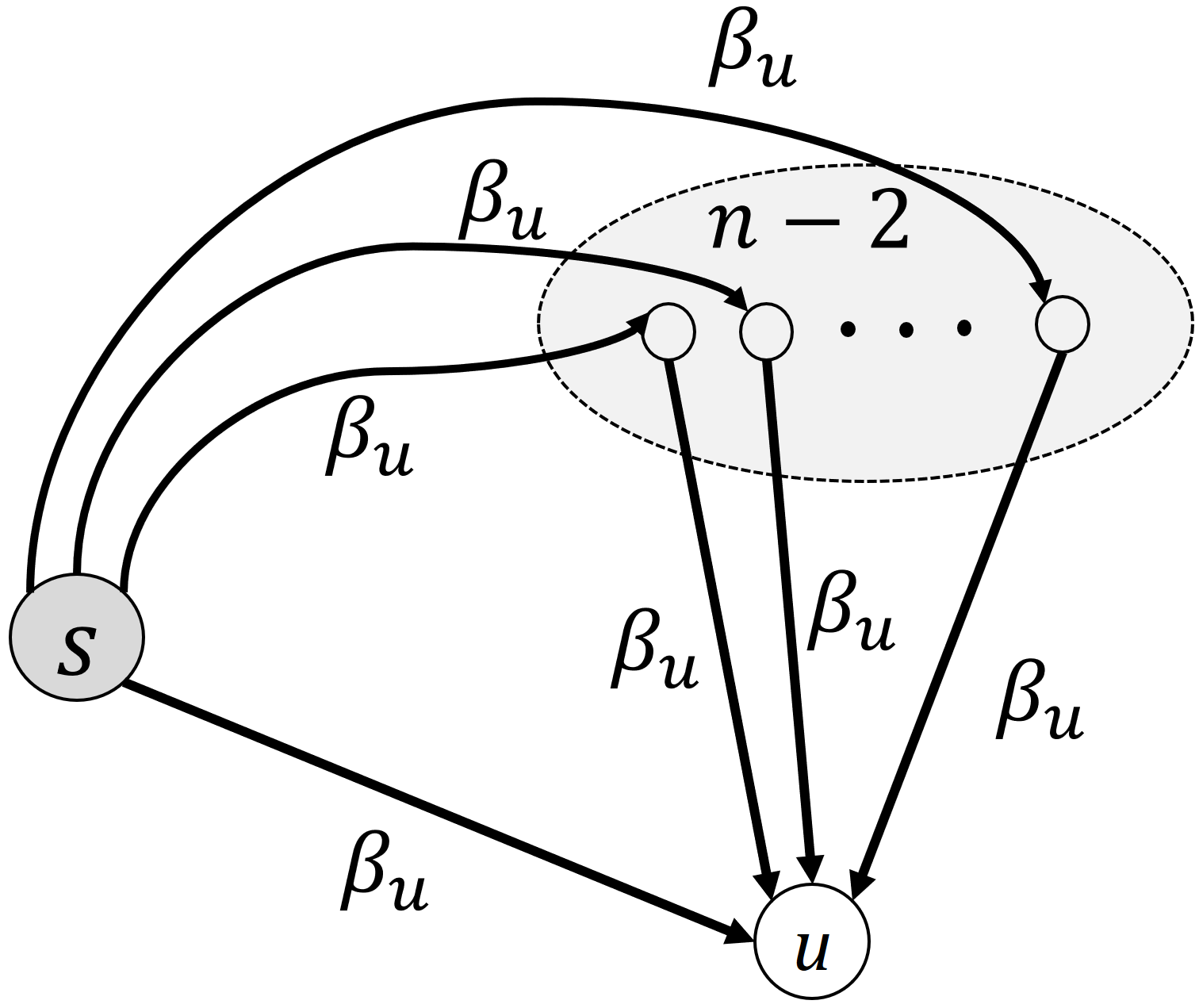}
\caption{Possible infection pathways in $n$ employees of which one of them ($s$) is infected.}
\label{Prob_infected}
\end{figure}

For simplicity, we hereafter use $P_I(\Delta t)$ to refer to the probability function $P_I(\Delta t, n, \beta_u,\kappa)$ for known values of $n$ and $\kappa$. Thus, computing the probability of infection per employee leads directly to the expected number of infected employees. Let us denote the source of infection at $t=0$ by $s$, and let $u$ be an employee who stays healthy until $\Delta t-1$. There are then two ways $u$ can be infected on day $\Delta t$; via some direct contact with $s$, or via contact with one of the $n-2$ other employees. The transmission probability for each of the contacts is $\beta_u$, and the probability of infection for the source is one, while the probability of infection for the other employees is $P_I(\Delta t -1)$ (Figure \ref{Prob_infected}). This means that the probability that an arbitrary employee like $u$ stays healthy after $c$ contacts with $s$ is $(1-1\times \beta_u)^c$. Similarly, the probability that $u$ remains healthy after $c$ contacts with the other $n-2$ employees on day $\Delta t$ is $(1-P_I(\Delta t -1) \times \beta_u)^c$. These relationships imply that the infection probability can be calculated as a recursive function. 
Since we assume $\kappa$ contacts for each employee per day and all of them have an equal chance to occur, this function can be written as

\begin{equation}
P_I(\Delta t)=1-(1-P_I(\Delta t -1)) \times \left[(1-\beta_u)^{\kappa \frac{1}{n-1}} \times (1-P_I(\Delta t-1)\beta_u)^{\kappa (1-\frac{1}{n-1})} \right],
\label{Rec_n_I}
\end{equation}


where $\kappa \frac{1}{n-1}$ is the expected number of contacts between employee $u$ and the infected source employee $s$, and $\kappa (1-\frac{1}{n-1})$ is the expected number of contacts between employee $u$ and the other employees except $s$. Note that, $(1-P_I(\Delta t -1))$ is the probability that the employee stays healthy until day $\Delta t-1$.

Eq. \ref{Rec_n_I} computes the probability that an arbitrary employee gets infected after $t=\Delta t$ days when the employees have $\kappa$ contacts per day with full workplace occupancy (i.e., $occup=1$). For $0 \leq occup \leq 1$, all employees (including $s$) stay at home (and healthy) with probability $1-occup$. So the probability of one contact between $u$ and $s$ occurring is $occup^2$. Therefore, the infection probability formula for known $n$, $\kappa$ and $occup$ can be extended as follows

\begin{dmath}
P_I(\Delta t)=1-(1-P_I(\Delta t -1))\left[(1-\beta_u)^{occup^2 \times \kappa \frac{1}{n-1}} \times (1-P_I(\Delta t-1)\beta_u)^{occup \times \kappa (1-\frac{1}{n-1})} \right].
\label{Rec_occuop_n_I}
\end{dmath}
Also, we set $P_I(0)=0$ without loss of generality as the scenario we assume here.
After computing the probability of infection, the expected number of infected employees can be straightforwardly computed by applying Eq. \ref{Exp_n_I}. 

\begin{figure}[t]
\centering
\includegraphics[width=12cm]{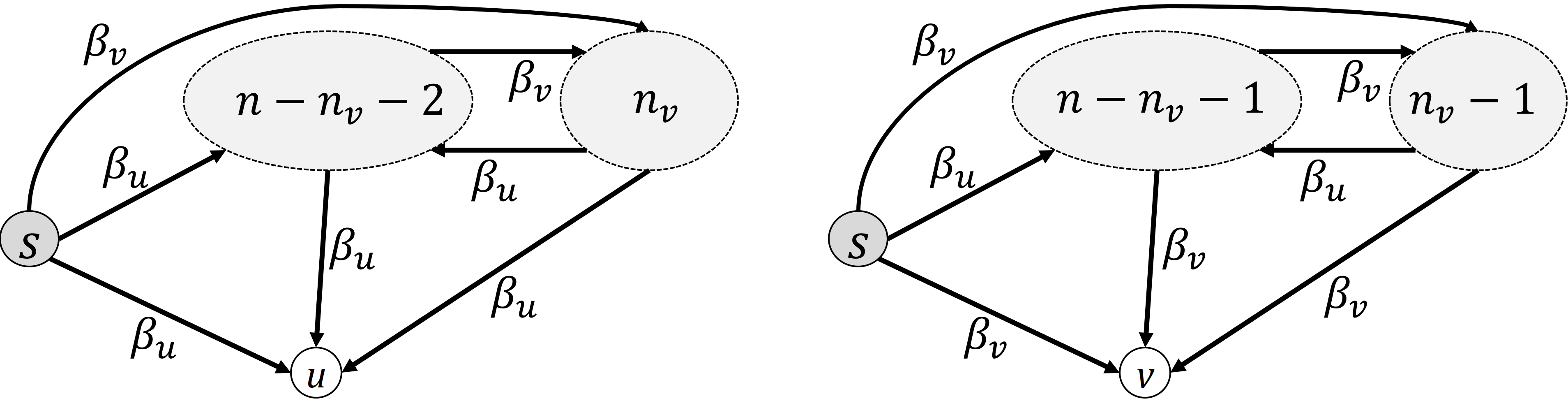}
\caption{Infection possibilities of $n$ employees of which one of them ($s$) is infected and there are two groups of vaccinated and unvaccinated employees. The left subfigure shows the scenario for an unvaccinated employee $u$ and the right one shows the scenario for a vaccinated employee $v$.}
\label{Prob_infected_v}
\end{figure}

Now, let us extend the above computation to two different groups, namely vaccinated and unvaccinated employees. We assume the disease transmission probability for the vaccinated group is $\beta_v ~(\beta_v \ll \beta_u)$. Without loss of generality, we consider the source of infection $s$ as an unvaccinated employee and compute the probability of infection in two cases: (\textit{i}) for an unvaccinated employee $u$, and (\textit{ii}) for a vaccinated employee $v$ (see Fig. \ref {Prob_infected_v}). There are three ways of infecting $u$: (\textit{i}) via a direct contact with $s$, (\textit{ii}) via contact with one of the $n_v$ vaccinated employees, or (\textit{iii}) via contact with one of the $n-n_v-2$ unvaccinated employees. The transmission probability for each of these contacts is $\beta_u$ because $u$ is an unvaccinated employee, however, the transmission probability for the vaccinated group is $\beta_v$. Let $P^u_I(\Delta t -1)$ and $P^v_I(\Delta t -1)$ be the probability of infection for an unvaccinated employee and a vaccinated employee, respectively. Thus, the recursive equations for each group can be written as

\begin{dmath}
P^u_I(\Delta t)=1-(1-P^u_I(\Delta t-1))\left[(1-\beta_u)^{occup^2 \times \kappa \frac{1}{n-1}} \times 
\left(1- P^u_I(\Delta t-1) \beta_u \right)^{occup \times \frac{n-n_v-1}{n-1} \times \kappa (1-\frac{1}{n-1})} \times
\left(1- P^v_I(\Delta t-1)\beta_v \right)^{occup \times \frac{n_v}{n-1} \times \kappa(1-\frac{1}{n-1})} 
\right],
\label{U_Rec_occuop_n_I}
\end{dmath}

and 

\begin{dmath}
P^v_I(\Delta t)=1-(1-P^v_I(\Delta t-1))\left[(1-\beta_v)^{occup^2 \times \kappa \frac{1}{n-1}} \times 
\left(1- P^u_I(\Delta t-1) \beta_v \right)^{occup \times \frac{n-n_v-1}{n-1} \times \kappa (1-\frac{1}{n-1})} \times
\left(1- P^v_I(\Delta t-1)\beta_v \right)^{occup \times \frac{n_v}{n-1} \times \kappa(1-\frac{1}{n-1})} 
\right],
\label{V_Rec_occuop_n_I}
\end{dmath}

where $P^u_I(0)=P^v_I(0)=0$. 
One advantage of these recursive formulas is that $P^u_I(\Delta t)$ and $P^v_I(\Delta t)$ can be efficiently computed in linear time to $\Delta t$ using a simple bottom-up approach. Finally, the expected number of infected employees $\Delta t$ days after an  infection is introduced can be computed as

\begin{equation}
\mathbb{E}\left[ n_I(\Delta t, n,n_v,\beta_u,\beta_v,\kappa,occup) \right]=1+ (n-n_v-1)\times P^u_I(\Delta t)+n_v \times P^v_I(\Delta t).
\label{Vaccin_Exp_n_I}
\end{equation}

Thus, Eq. \ref{Vaccin_Exp_n_I} provides a recursive formula to compute the number of infections over time. An implementation of this equation and comparison with simulation results is presented in the Appendix.


\subsection{Computing optimal presence rate}
As explained, the expected number of infected employees $\Delta t$ days after an infection arrives at the facility can be computed using Eq.\ref{Vaccin_Exp_n_I}. For simplicity, we denote this expected number by $\mathbb{E}\left[ n_I(\Delta t)\right]$. Let $\rho$ be the per day probability that an infection arrives at the facility. This probability can be determined using two straightforward approaches. First, using the (recent) historical data of incidence in the company, and second, using the local incidences and applying the ratio of the vaccination rate in the neighborhood of the facility to the vaccination rate in the company. Therefore, the cumulative number of infected employees for a time interval $T$ days can be computed as follows

\begin{equation}
\mathbb{E}(n_I)=Accum_I(T,\rho)=\rho \times Z(T),
\label{Accum_infect}
\end{equation}
where $Z(T)$ is defined

\begin{dmath}
Z(T) = \left\{ \begin{array}{rcl}
\mathbb{E}\left[ n_I(0)\right]+\mathbb{E}\left[ n_I(1)\right]-\frac{\mathbb{E}\left[ n_I(1)\right] \times \mathbb{E}\left[ n_I(0)\right]}{n} & \mbox{, if}
& T=1 \\ \mathbb{E}\left[ n_I(T)\right]+Z(T-1)-\frac{\mathbb{E}\left[ n_I(1)\right] \times Z(T-1)}{n} & \mbox{, if } & T  \geq 2
\end{array}\right.
\end{dmath}

Note that $Z(T)$ is a linear recursive formula to compute the cumulative number of infections for $t=0,1,\dots, T$ by removing the expected overlapping infected employees over time.

Eq. \ref{Accum_infect} computes the total number of infected employees over a time interval $T$ days. If employees are tested every $\tau$ days on average, and they uniformly distribute in the test interval (having $k=\frac{n}{\tau}$ tests per day on average), the expected time to detect the infection can be computed as

\begin{dmath}
\overline {\tau}=\left [1-{(1-Pr(t))}^{k} \right ]+ \sum_{t=2}^{\tau} \left[ t \times (1-{(1-Pr(t))}^{k}) \times \prod_{t'=1}^{t-1}  {(1-Pr(t'))}^{k} \right],
\label {exp_detect}
\end{dmath}

where $Pr(t)=\frac{\mathbb{E}\left[ n_I(t)\right]}{n}$ is the probability of infection per employee $t$ days after an infection arrives at the facility, so, ${1-(1-Pr(t))}^{k}$ is the probability of detecting at least one infected employee by testing a subgroup of $k$ after $t$ days. For detecting the infection $t>1$ days after of arrival, we must compute the probability that it has been not detected in the days before $t$, i.e. $t'<t$, which is computed using $\prod_{t'=1}^{t-1}  {(1-Pr(t'))}^{k}$. 

$\overline {\tau}$ in Eq.\ref{exp_detect} is the expected time to detect an infection if the employees are tested every $\tau$ days on average. Therefore, the expected number of infected employees before detecting an infection is obtained by $\mathbb{E}\left[ n_I(\tau,n,n_v,\beta_u,\beta_v,\kappa,occup) \right]$. So, the total productivity (see Eq.\ref{prod_obj}) of a facility can be determined for any given $occup$ in $[0,1]$, and as a function of the other parameters.

\subsection{A practical multi-objective solution approach}
As previously described, the optimal workplace presence strategy is formulated in the framework of a bi-objective optimization problem as follows

\begin{equation}
\begin{aligned}
&Minimize~~~\mathbb{E}(n_I),\\
&Maximize~~~Total~Productivity,\\
&~~~~Subject~to:\\
&~~~~~~~~~~~~occup \ge An~Occupancy~Threshold.\\
\label{bi_obj_model}
\end{aligned}
\end{equation}

\begin{figure}[t]
\begin{subfigure}{.5\textwidth}
  \centering
  \includegraphics[width=1.0 \linewidth]{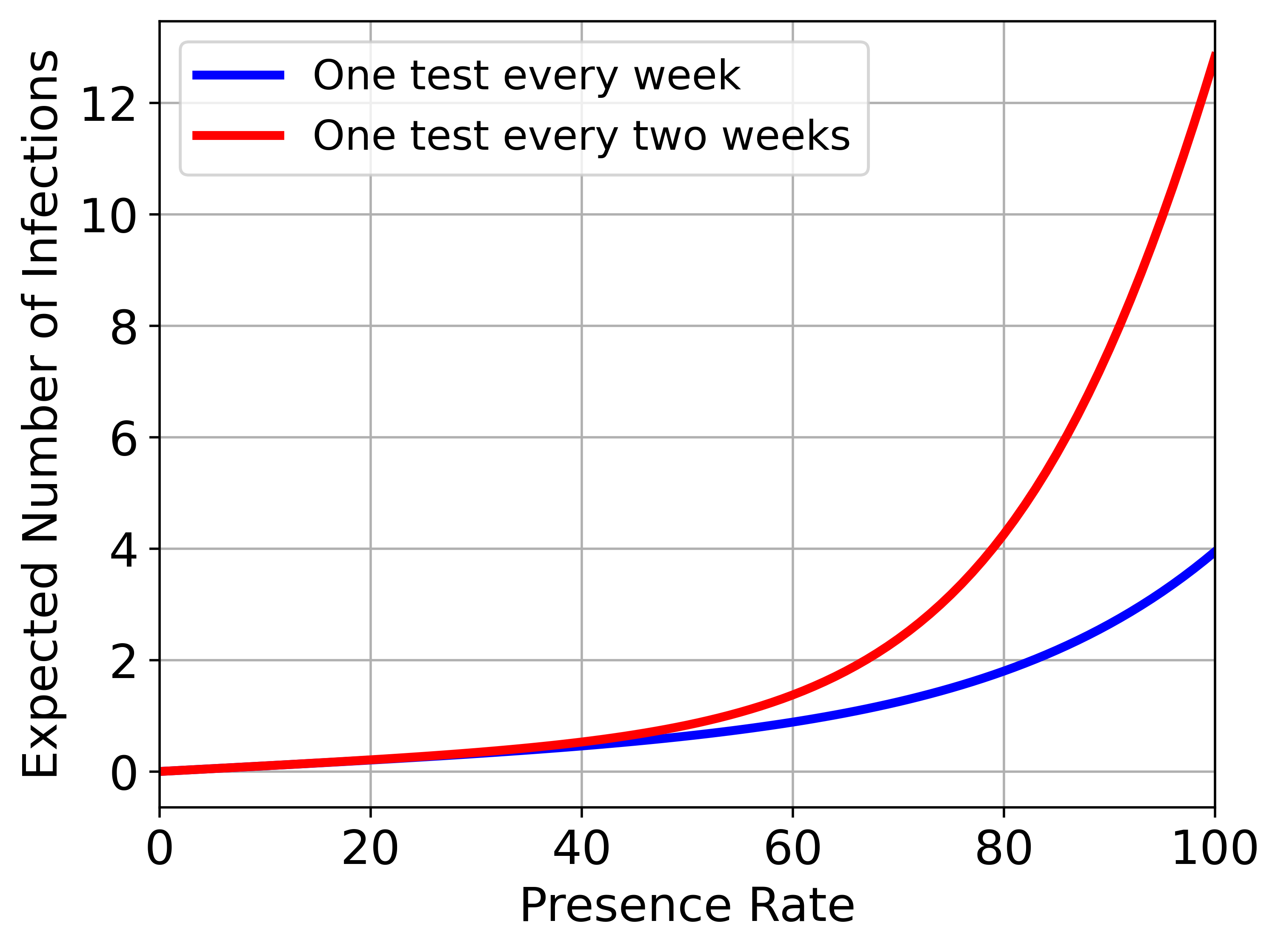}  
\end{subfigure}
\begin{subfigure}{.5\textwidth}
  \centering
  \includegraphics[width=1.0 \linewidth]{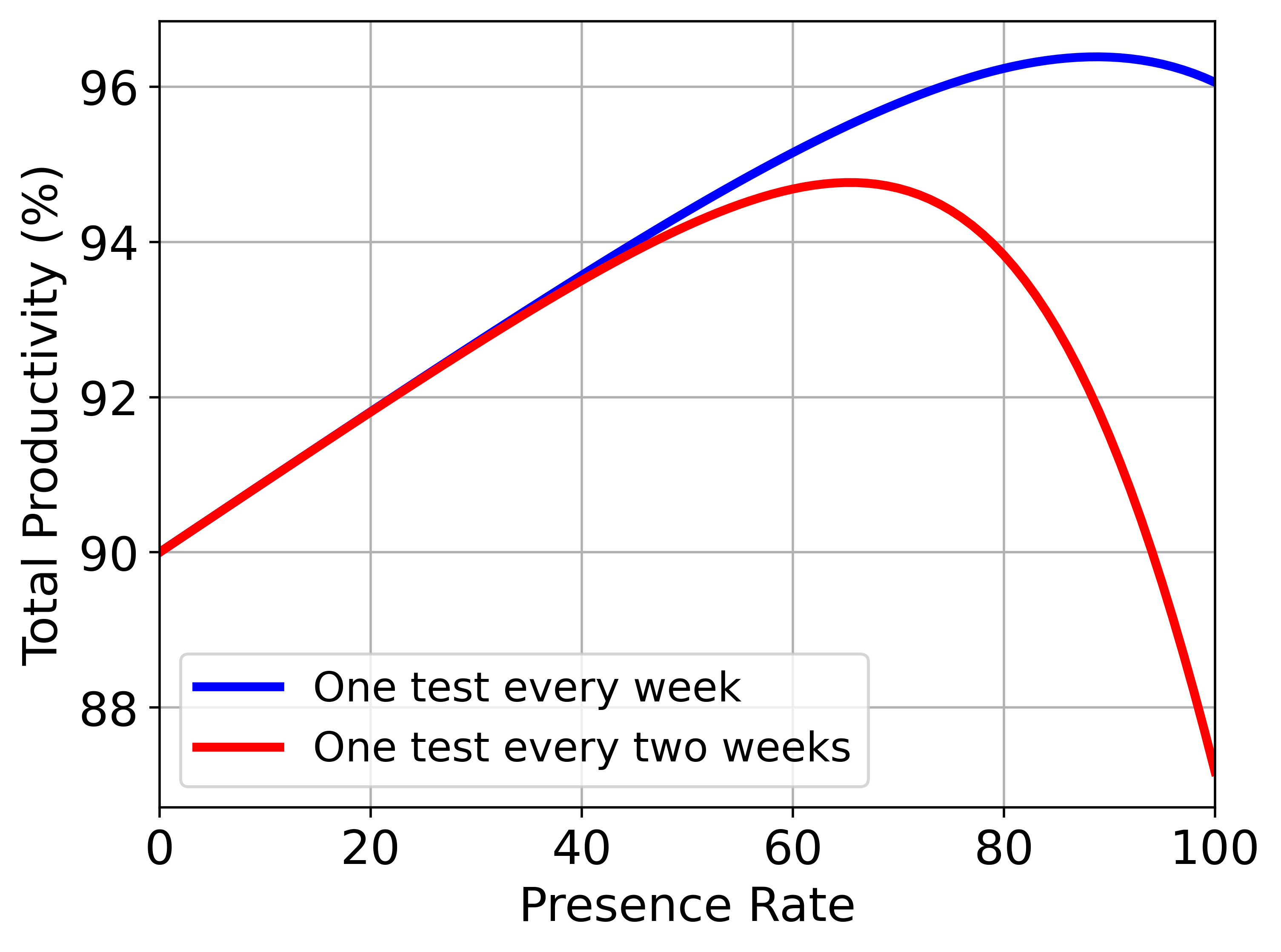}  
\end{subfigure}
\caption{Effect of occupancy on the expected number of infections (left panel) and on the total productivity (right panel) for one test per week and one test per two weeks. In these simulations, we assumed an organization with 100 employees of which half are vaccinated, with 15\% contact rate, and the home productivity ratio 0.90.}
\label{trade_off_fig}
\end{figure}

The first objective is minimizing the expected number of infected employees in Eq. (\ref{Accum_infect}), and the second objective is maximizing the total productivity of the organization via Eq. (\ref{prod_obj}). The only constraint that we considered in the model is related to the minimum possible workplace occupancy, the \textit{Occupancy Threshold}, which is the minimum number of employees that must be present at the facility to perform tasks for which physical presence is required. It is also possible to include additional lower or upper bound constraints on the occupancy variable. The model has two conflicting objectives. Clearly, by increasing workplace occupancy, the expected number of infected employees will increase as well. This is, however, not the case for the second objective, because by increasing occupancy, the number of infected employees also increases, but total productivity will eventually decrease as the productivity of infected employees is zero. Figure \ref{trade_off_fig} shows the effect of occupancy on the first and second objective separately for the scenarios of one test per week (blue curve) and one test per two weeks (red curve). As expected, the number of infections is a strictly increasing function of occupancy rate, while, the productivity function is increasing for the low occupancy, and after reaching the maximum productivity (e.g., for $occup \approx 88\%$ in the first scenario and for $occup \approx 65\%$ in the second one), then becomes a decreasing function of occupancy.

The outcome of the optimization problem presented in \ref{bi_obj_model} is a set of trade-off, or Pareto optimal solutions, which are defined as solutions that are not improved for an objective unless it sacrifices the other objective \cite{coello2007evolutionary}. There are different approaches to solve multi-objective optimization problems, such as the weighted-sum method \cite{kim2005adaptive}, the lexicographic method \cite{rentmeesters1996theory}, the $\epsilon-constraint$ approach \cite{miettinen2012nonlinear}, goal programming \cite{chang2007multi}, or evolutionary algorithms \cite{deb2011multi, coello2007evolutionary}. All of these approaches have their advantages and disadvantages. Some of them have high time complexity, others work only for differentiable functions or convex models, while some provide no guarantee of finding all the Pareto-optimal solutions. We therefore propose a quick and practical method to compute all Pareto optimal solutions of the model \ref{bi_obj_model} efficiently.

The method utilizes some observations to solve the model. First, $occup$ is a continuous variable, but in reality, there are at most $n$ possible choices for the number of employees present in the facility. So initially, for any $occup \in \{ 0, \frac{1}{n}, \frac{2}{n}, \dots, \frac{n}{n}\}$ which satisfy the occupancy threshold, we compute the expected detection time based on Eq. (\ref{exp_detect}), and the expected number of infected employees and its corresponding total productivity based on Eq. (\ref{Vaccin_Exp_n_I}) and Eq.(\ref{prod_obj}), respectively. All of these steps can be handled in $O(n\tau)$ time using the presented linear recursive equations. We denote this solution set by $S$. Now, we can find the \textit{non-dominated} solutions of $S$ and report them as the Pareto-optimal solutions of the problem. A solution $\bar{s}\in S$ is a non-dominated solution if there is no other solution in $S$ such that it is better than $\bar{s}$ in both objectives, i.e., its expected number of infections is less than the expected number of infections of $\bar{s}$, and simultaneously, its productivity is more than the productivity of $\bar{s}$. Since the solutions of $S$ are constructed one by one, from the minimum expected infections to the maximum one, the Pareto-optimal solutions of the problem can be computed in linear time using a \textit{sweep-line} approach \cite{jensen2003reducing}. The pseudocode of the proposed algorithm is presented as follows.

\begin{algorithm}
\caption{Computing Pareto Optimal Workplace Presence Strategies}
\hspace*{\algorithmicindent} \textbf{Input:} Organizations's Parameters (the notation mentioned in Table \ref{table_params})\\
	\hspace*{\algorithmicindent} \textbf{Output:} All Pareto Optimal Strategies
\begin{algorithmic}
\STATE $max\_prod \leftarrow -1$
\STATE $i \leftarrow 0$
\WHILE{$i \le n$}
\STATE{$occup \leftarrow \frac{i}{n}$}
\IF{$occup \ge Occupancy\_Threshold$}
\STATE Compute $\overline{\tau}$ using Eq. (\ref{Accum_infect})
\STATE Compute expected number of infections using Eq. (\ref{Vaccin_Exp_n_I})
\STATE Compute the total productivity using Eq. (\ref{prod_obj}) and denote it by $TP$
\IF{$TP > max\_prod$}
\STATE Report the current strategy as a Pareto optimal strategy
\STATE $max\_prod \leftarrow TP$
\ENDIF
\ENDIF
\STATE $i \leftarrow i+1$
\ENDWHILE
\end{algorithmic}
\end{algorithm}

\section{Simulation Results and Discussions}
In this section, we show some numerical results from the proposed model and method for computing Pareto optimal solutions that minimize the expected number of infections while maximizing productivity. Since there are several influential input parameters in the proposed model, there are too many possible combinations of them to consider exhaustively. We therefore briefly display the output of the model for some diverse set of possible inputs, and instead, we provide an online optimization tool based on the model to allow interested readers to consider all possible combinations of the model parameters as input and observe the output. A user-friendly version of this tool is available at \textit{where2test.de/optimization}.

In our simulations, we consider a mid-sized organization with $n=100$ employees and suppose the 7-day incidence rate per 100,000 population is 500. Thus, the average probability of infection per employee in a week will be $\frac{500}{100,000}$, and consequently, the probability that an infection arrives at the facility is $\rho=1-(1-\frac{500}{100000})^n$.
We specifically consider two scenarios based on the disease transmission rate for unvaccinated employees ($\beta_u$). First, we choose the baseline value as $\beta_u=0.04$ from a possible range of values reported in a previous study~\cite{lelieveld2020model}. This baseline scenario can be regarded as the situation where \textit{Delta} is the dominant SARS-CoV-2 variant. We then consider the situation where the transmissibility increases 2.5-fold (i.e. $\beta_u=0.1$), which is consistent with the recently identified \textit{Omicron} variant \cite{yang2021sars, world2022enhancing, lyngse2021sars}. Additionally, we set the transmission rate of vaccinated employees to $\beta_v=(1-0.80)\beta_u$, which implies 80\% immunity for vaccinated individuals \cite{polack2020safety,baden2020efficacy, emary2021efficacy, fiolet2021comparing}.

\begin{figure}[p]
\begin{subfigure}{.32\textwidth}
  \centering
  \includegraphics[width=1 \linewidth]{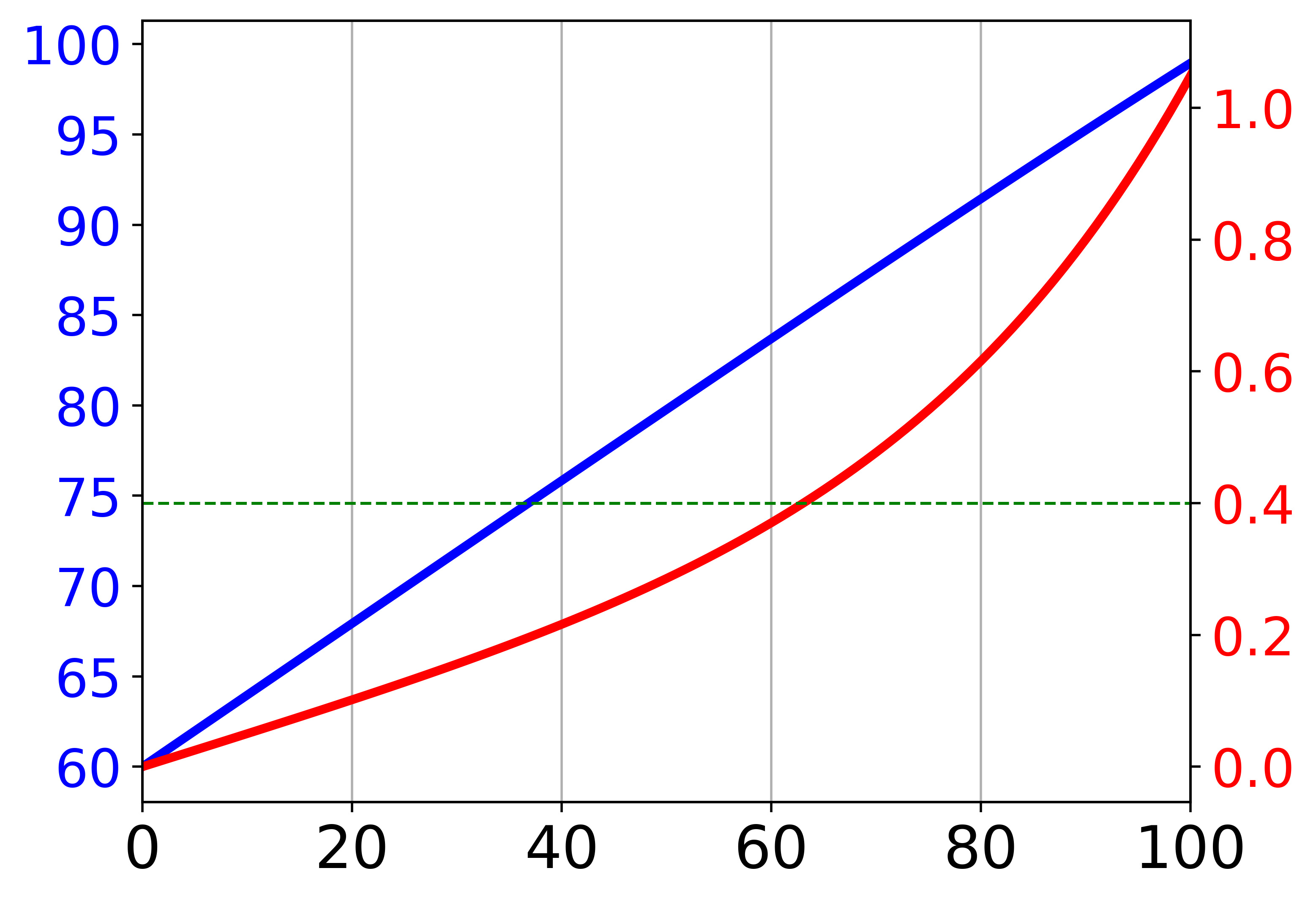} 
  \caption{}
\end{subfigure}
\begin{subfigure}{.32\textwidth}
  \centering
  \includegraphics[width=1 \linewidth]{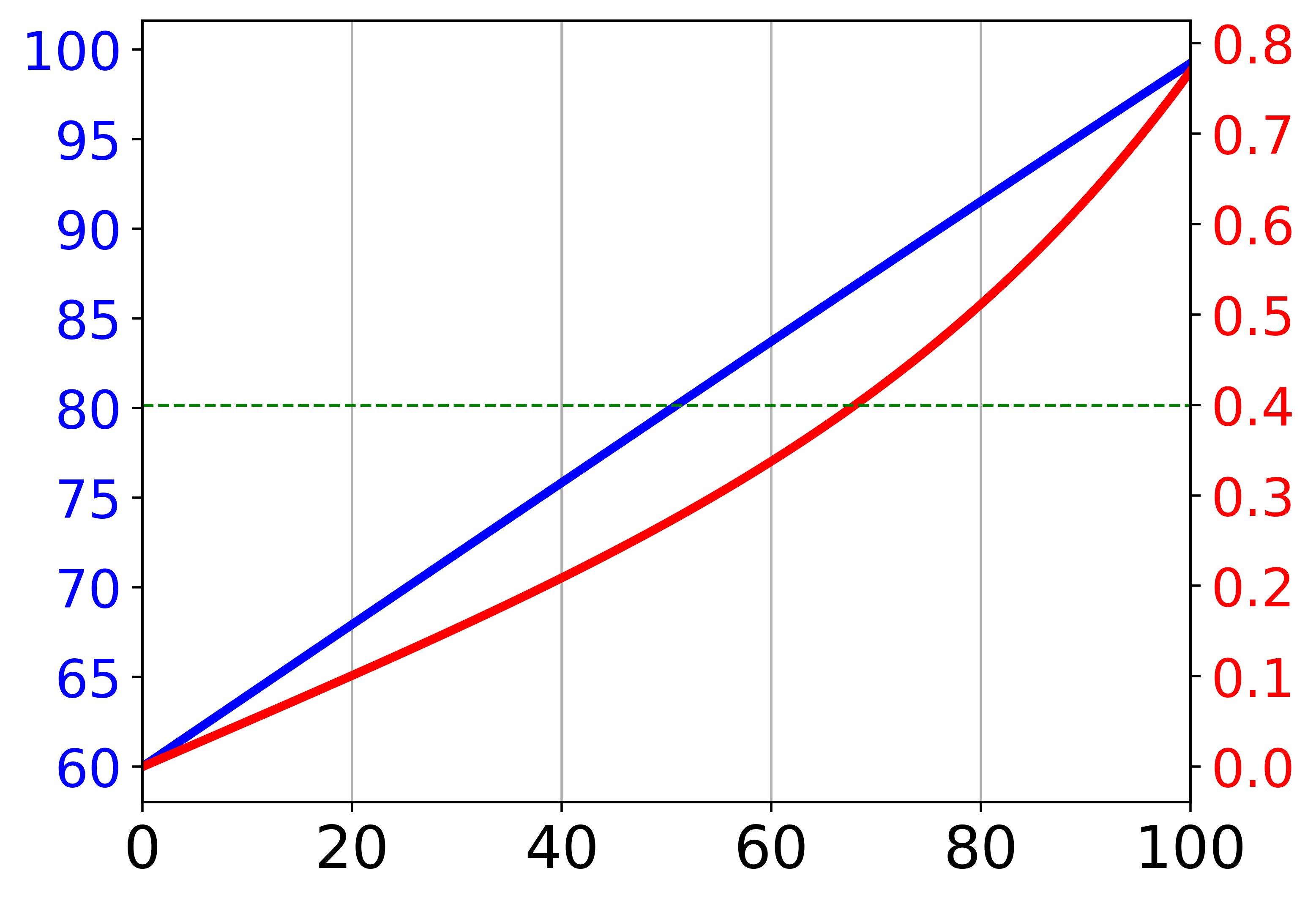} 
  \caption{}
\end{subfigure}
\begin{subfigure}{.32\textwidth}
  \centering
  \includegraphics[width=1 \linewidth]{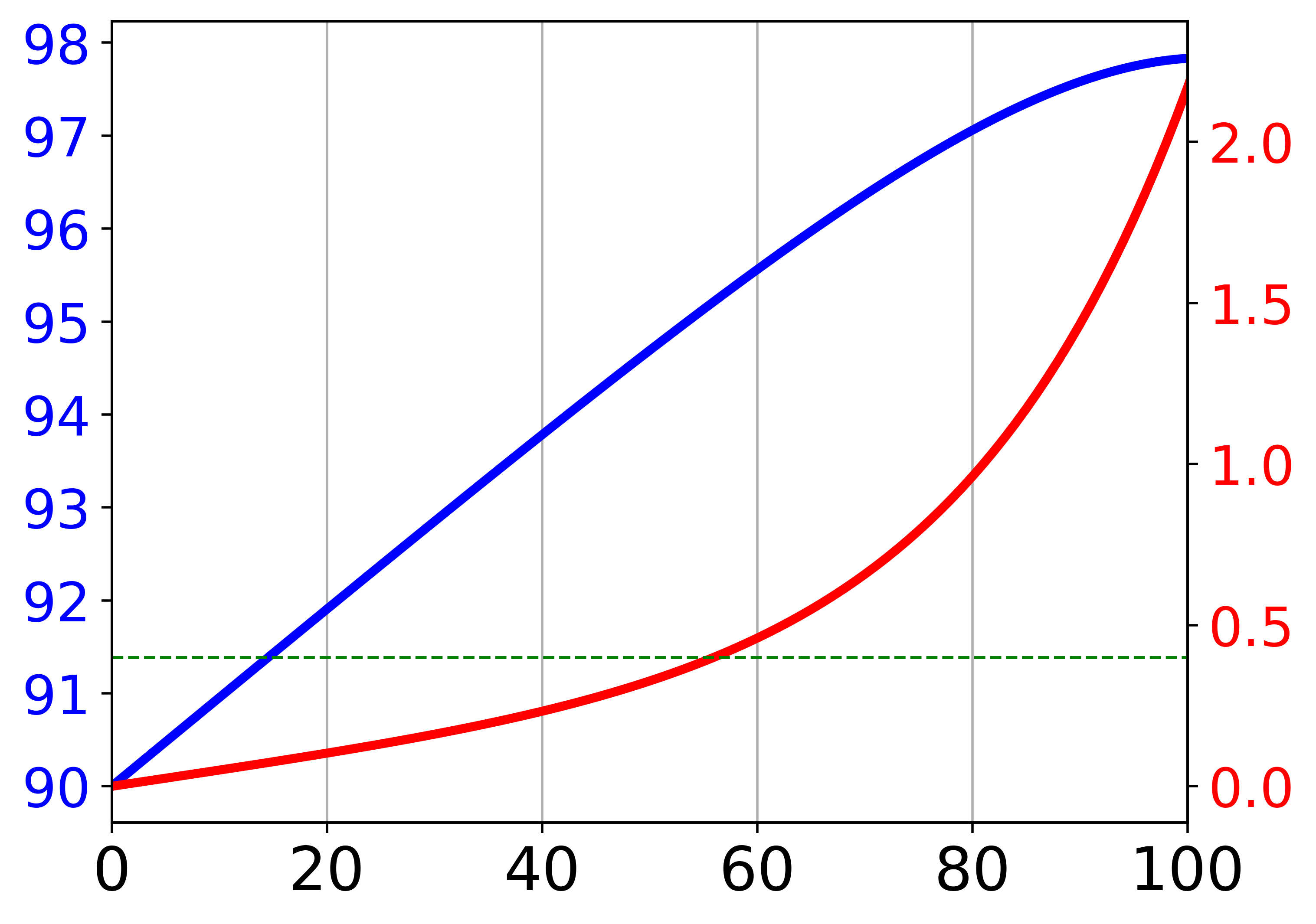} 
  \caption{}
\end{subfigure}

\begin{subfigure}{.32\textwidth}
  \centering
  \includegraphics[width=1 \linewidth]{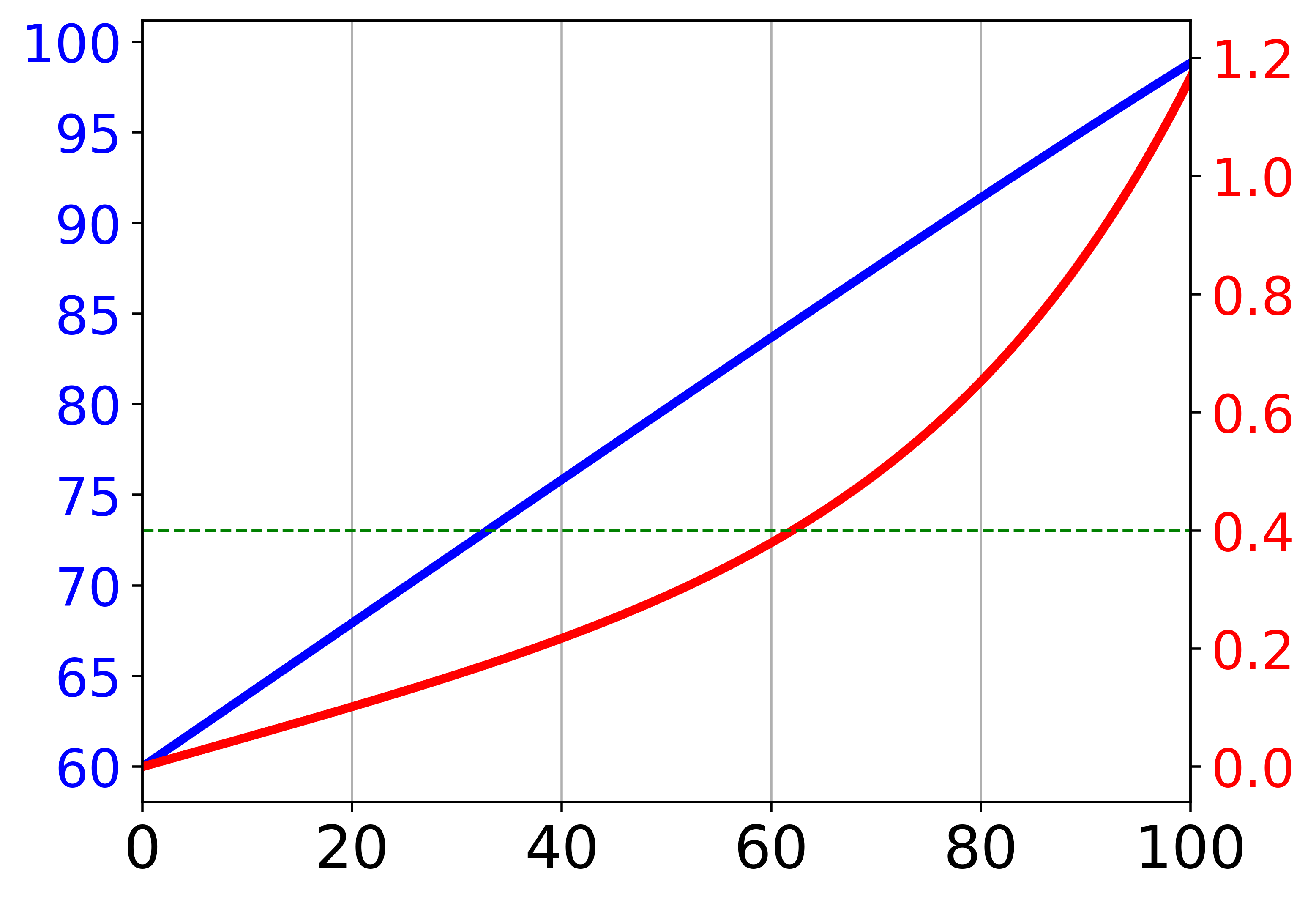} 
  \caption{}
\end{subfigure}
\begin{subfigure}{.32\textwidth}
  \centering
  \includegraphics[width=1 \linewidth]{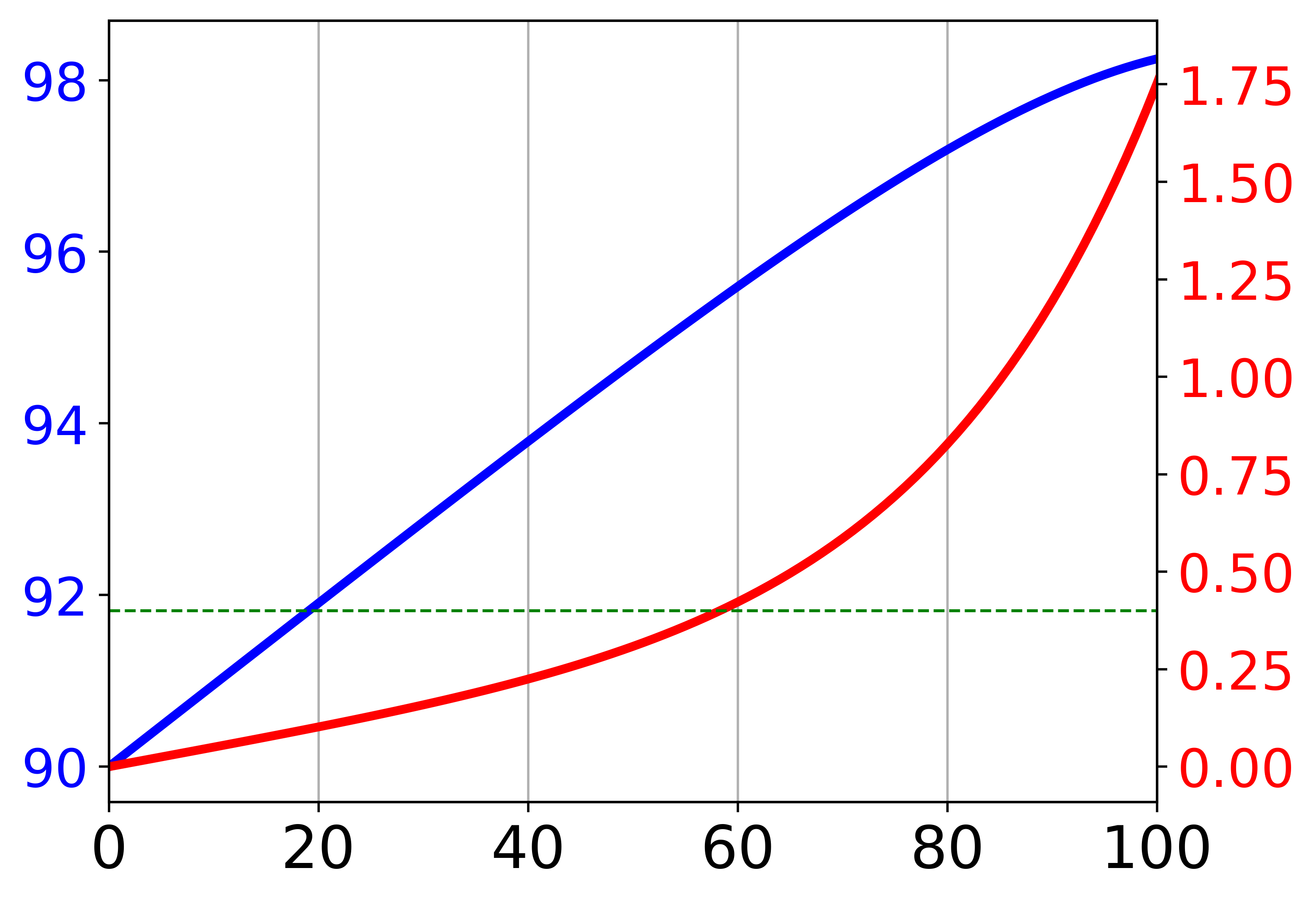} 
  \caption{}
\end{subfigure}
\begin{subfigure}{.32\textwidth}
  \centering
  \includegraphics[width=1 \linewidth]{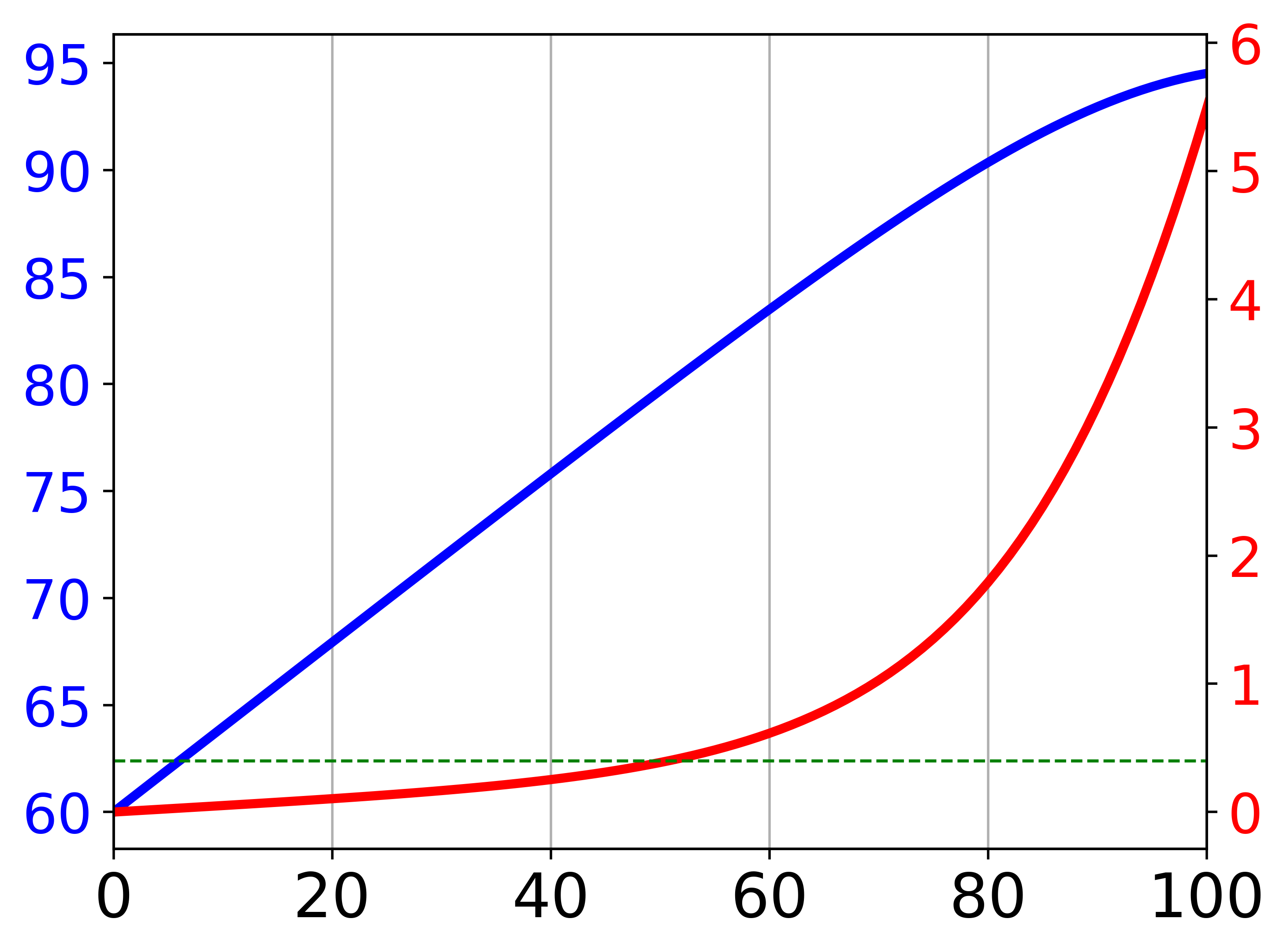} 
  \caption{}
\end{subfigure}

\begin{subfigure}{.32\textwidth}
  \centering
  \includegraphics[width=1 \linewidth]{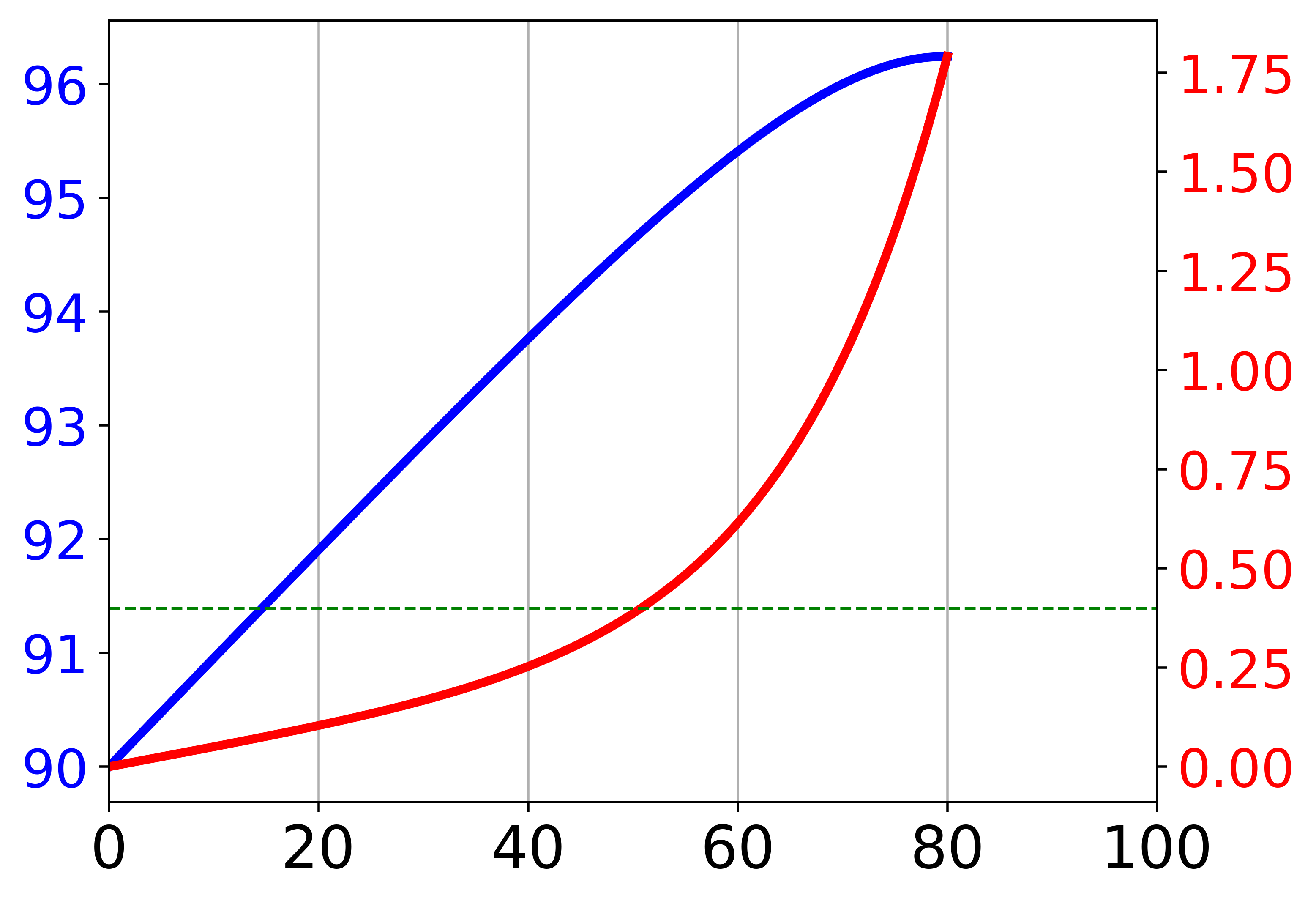} 
  \caption{}
\end{subfigure}
\begin{subfigure}{.32\textwidth}
  \centering
  \includegraphics[width=1 \linewidth]{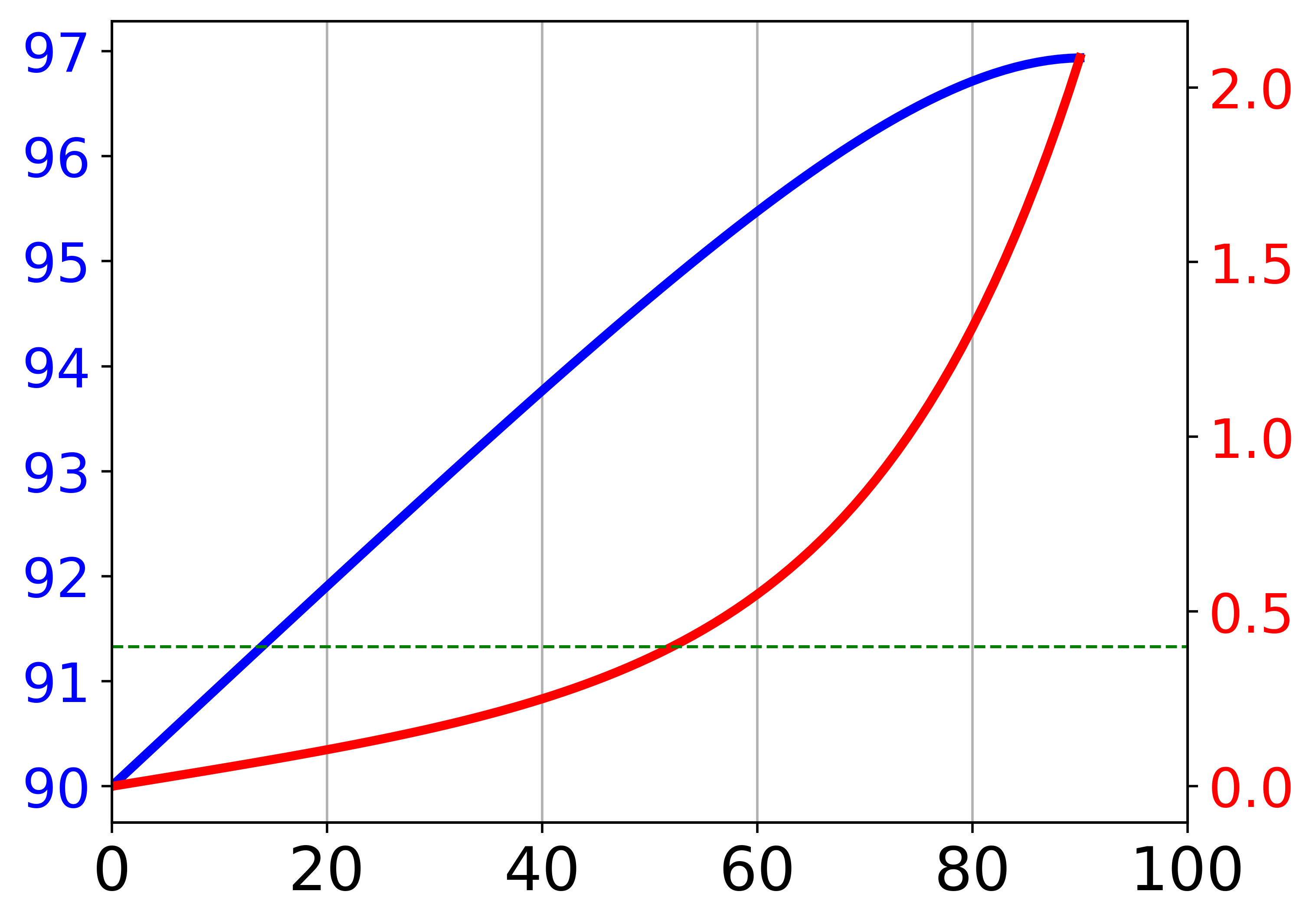} 
  \caption{}
\end{subfigure}
\begin{subfigure}{.32\textwidth}
  \centering
  \includegraphics[width=1 \linewidth]{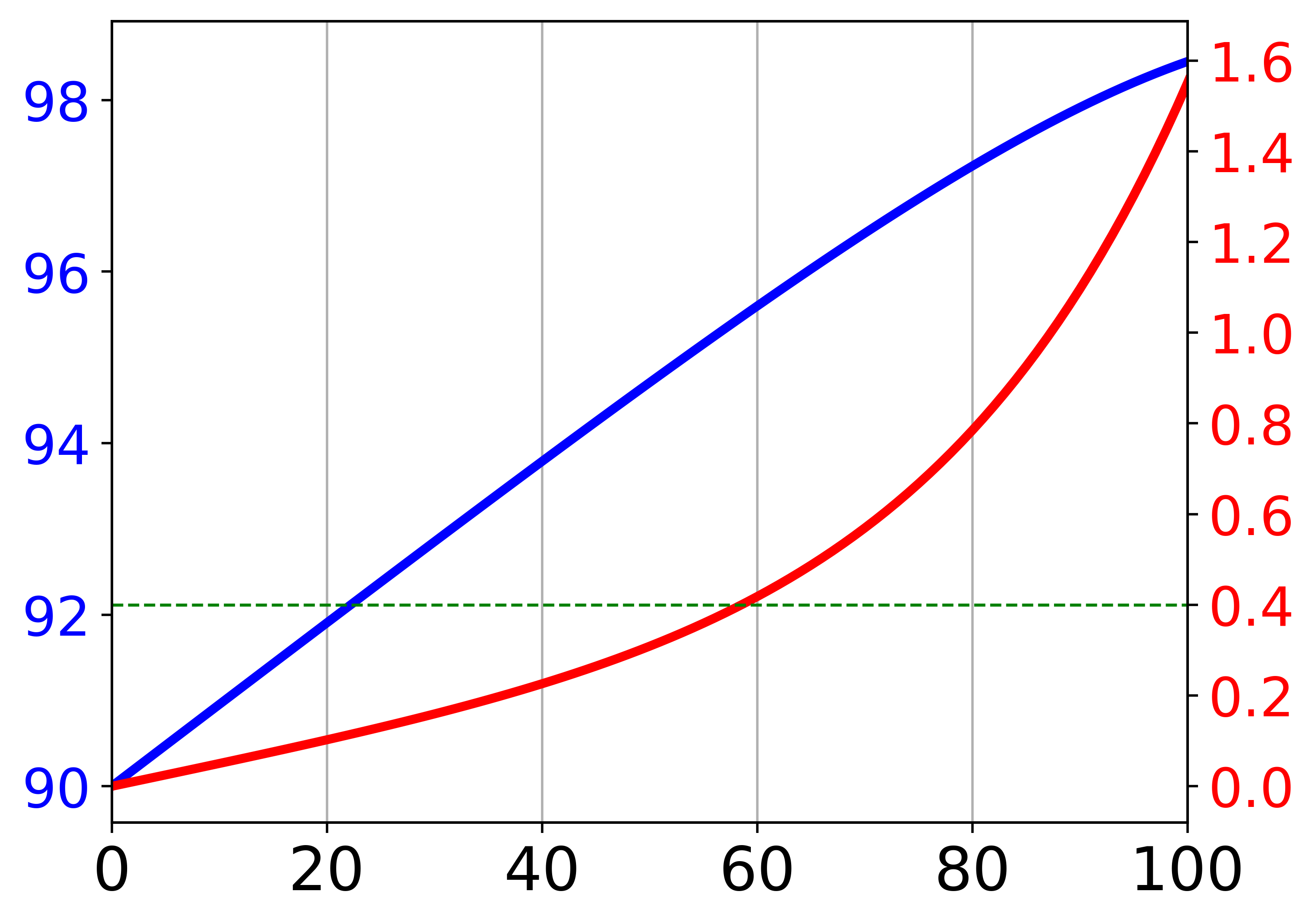} 
  \caption{}
\end{subfigure}

\begin{subfigure}{.32\textwidth}
  \centering
  \includegraphics[width=1 \linewidth]{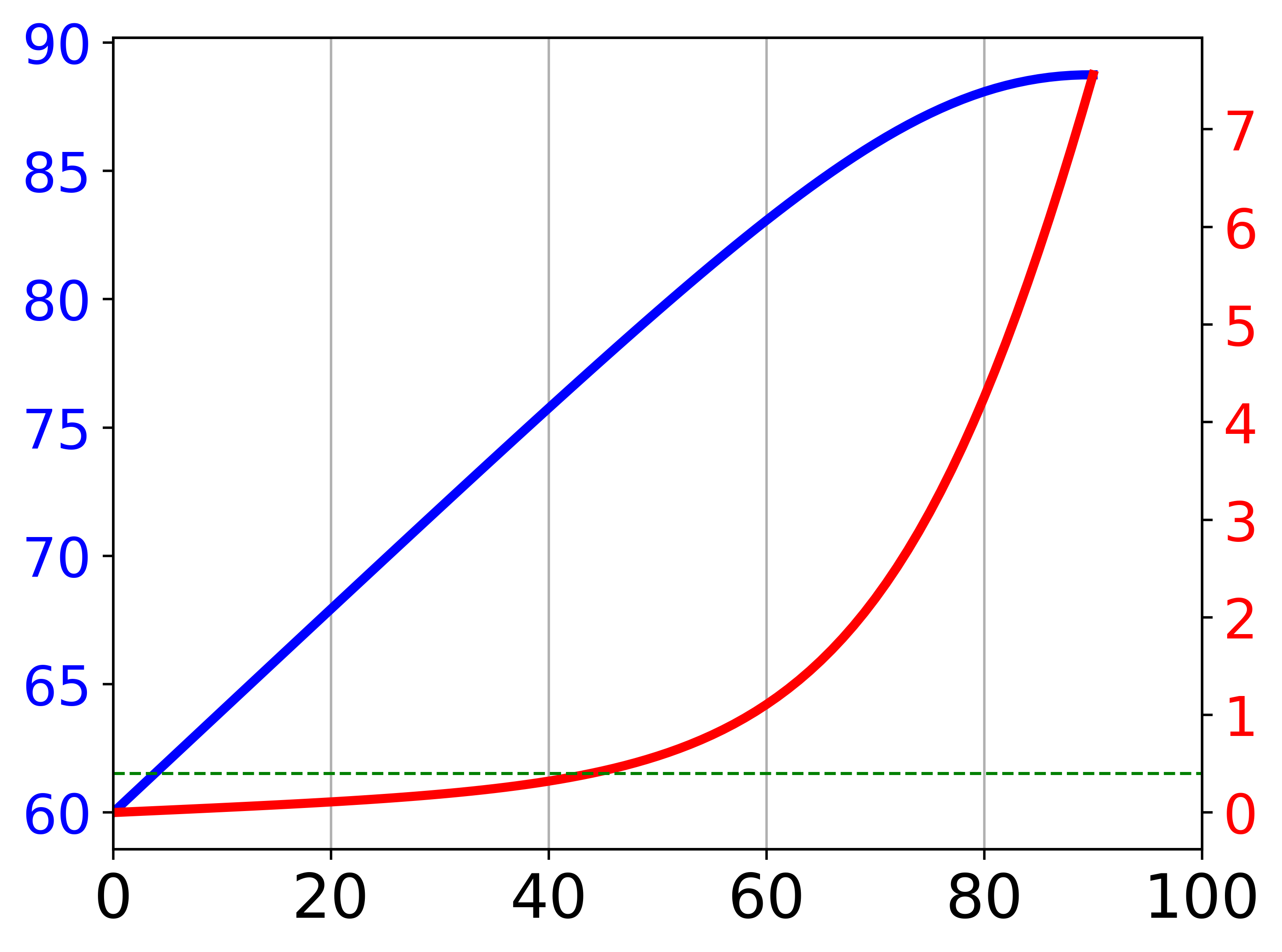} 
  \caption{}
\end{subfigure}
\begin{subfigure}{.32\textwidth}
  \centering
  \includegraphics[width=1 \linewidth]{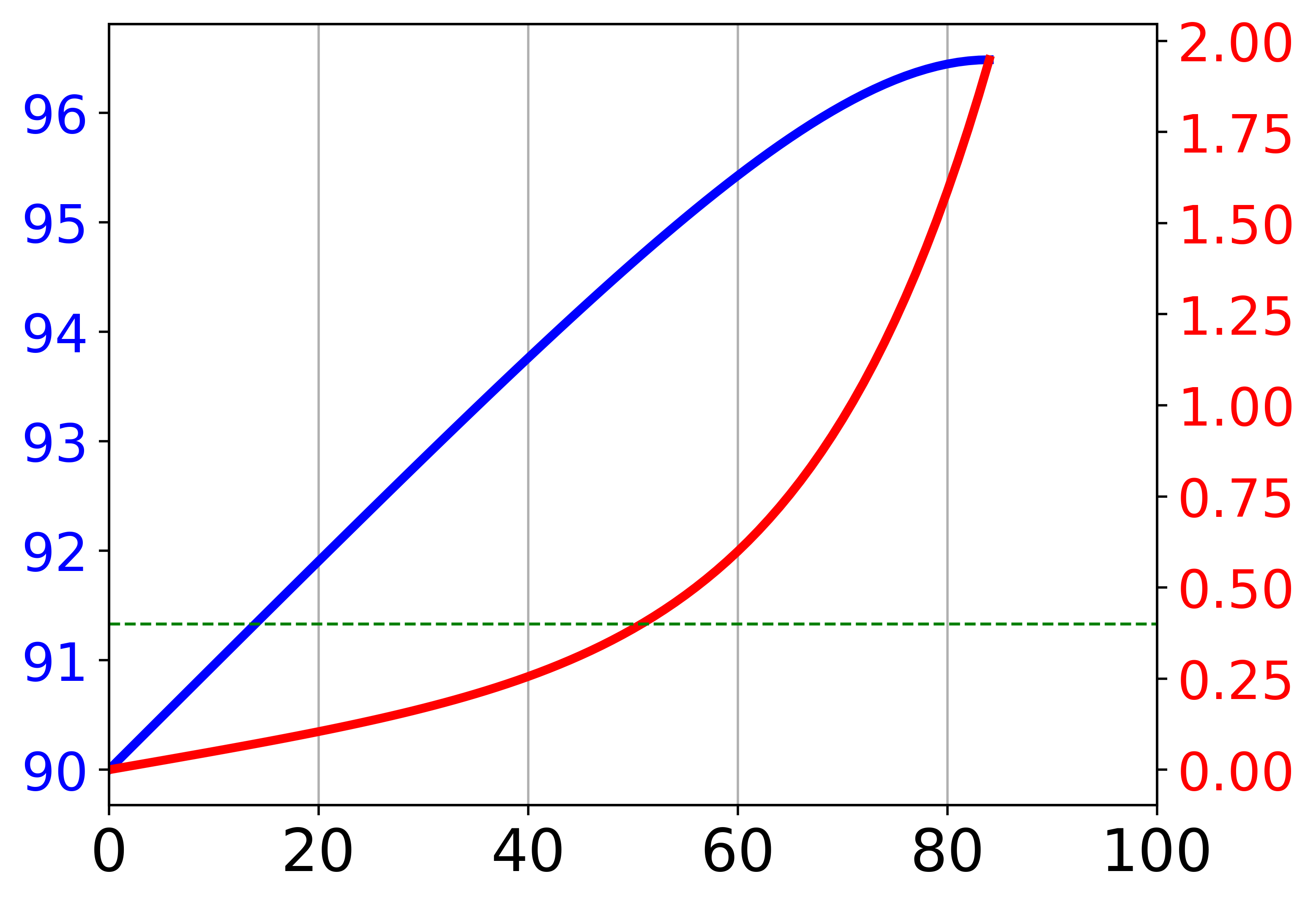} 
  \caption{}
\end{subfigure}
\begin{subfigure}{.32\textwidth}
  \centering
  \includegraphics[width=1 \linewidth]{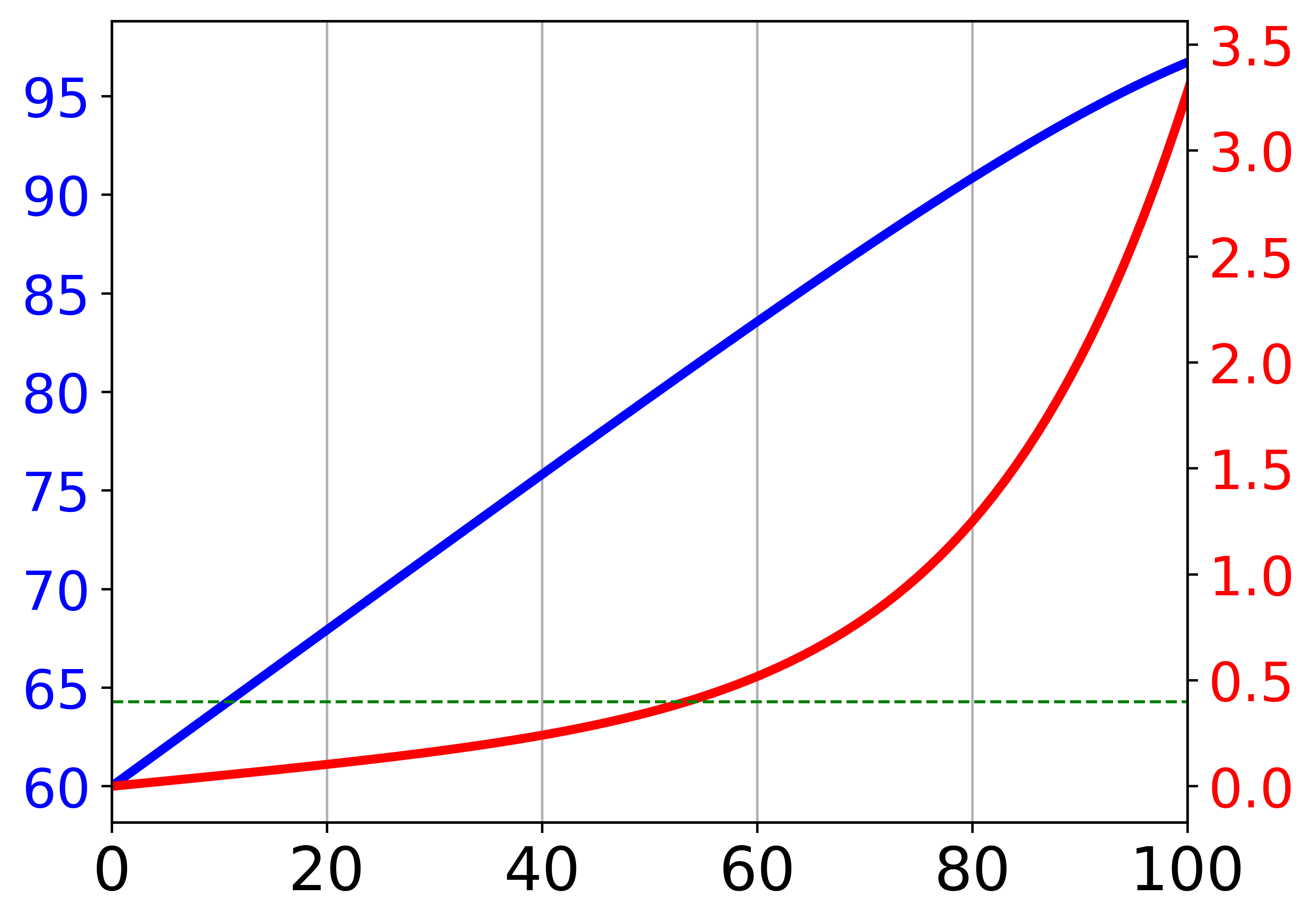} 
  \caption{}
\end{subfigure}

\begin{subfigure}{.32\textwidth}
  \centering
  \includegraphics[width=1 \linewidth]{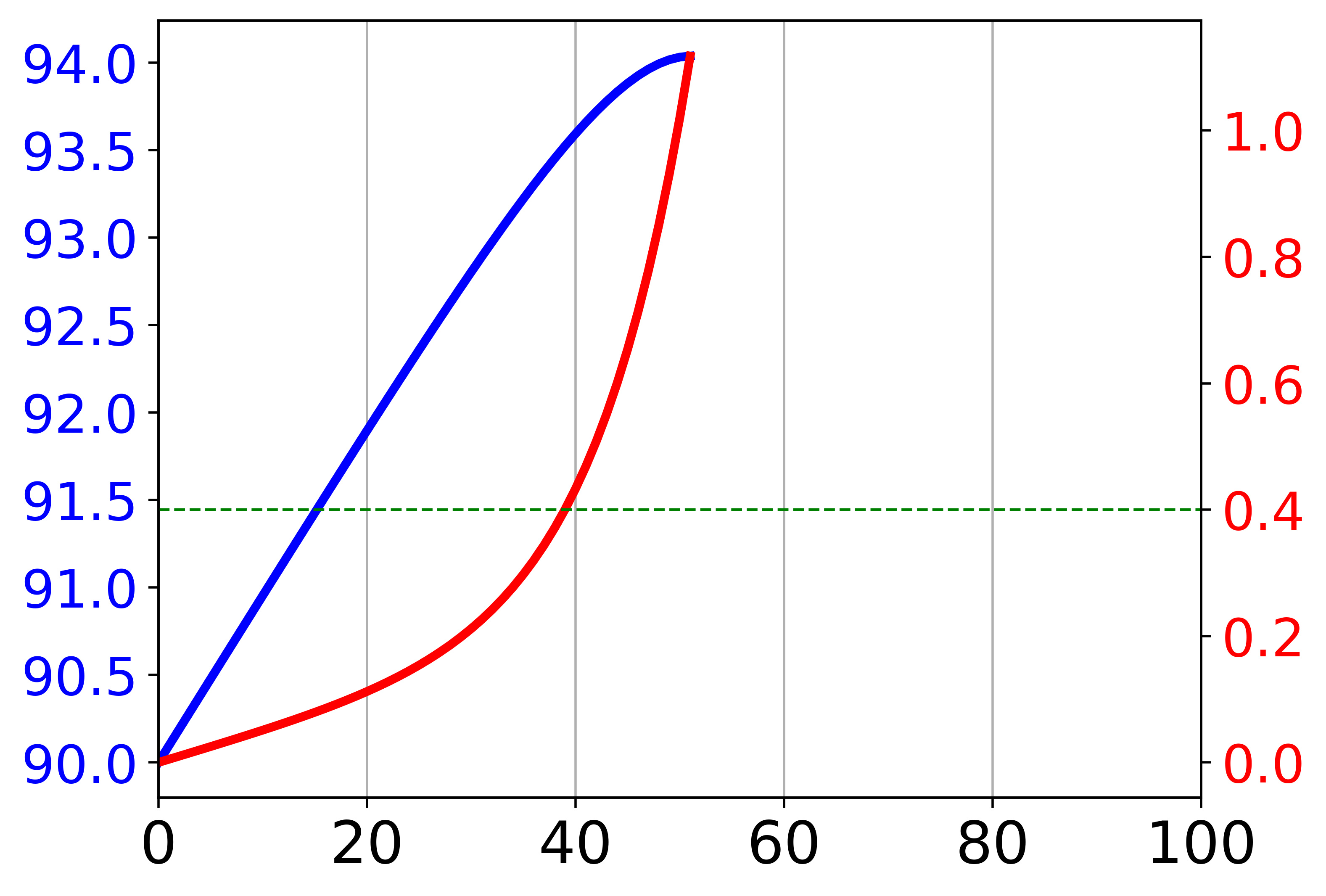} 
  \caption{}
\end{subfigure}
\begin{subfigure}{.32\textwidth}
  \centering
  \includegraphics[width=1 \linewidth]{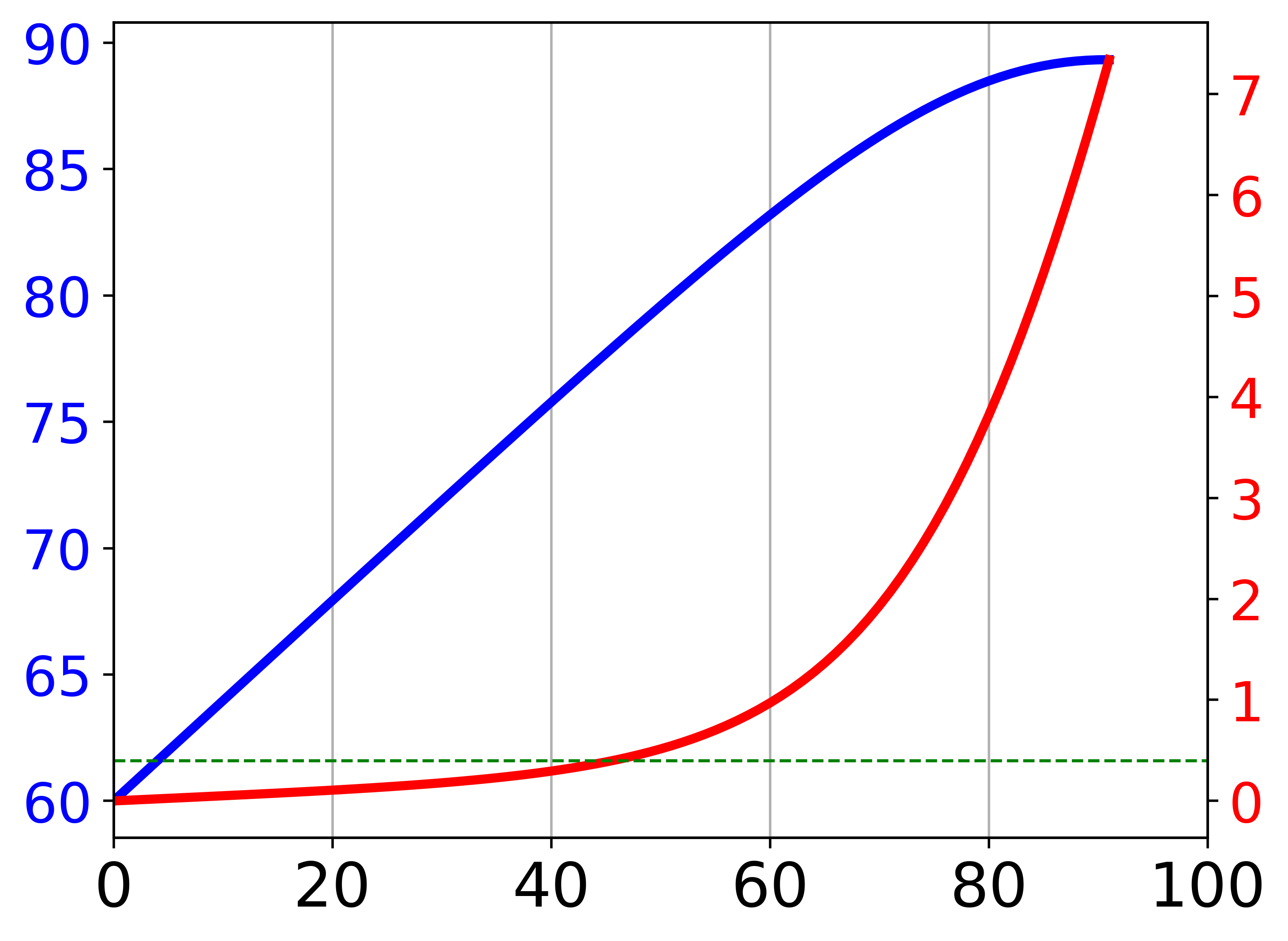} 
  \caption{}
\end{subfigure}
\begin{subfigure}{.32\textwidth}
  \centering
  \includegraphics[width=1 \linewidth]{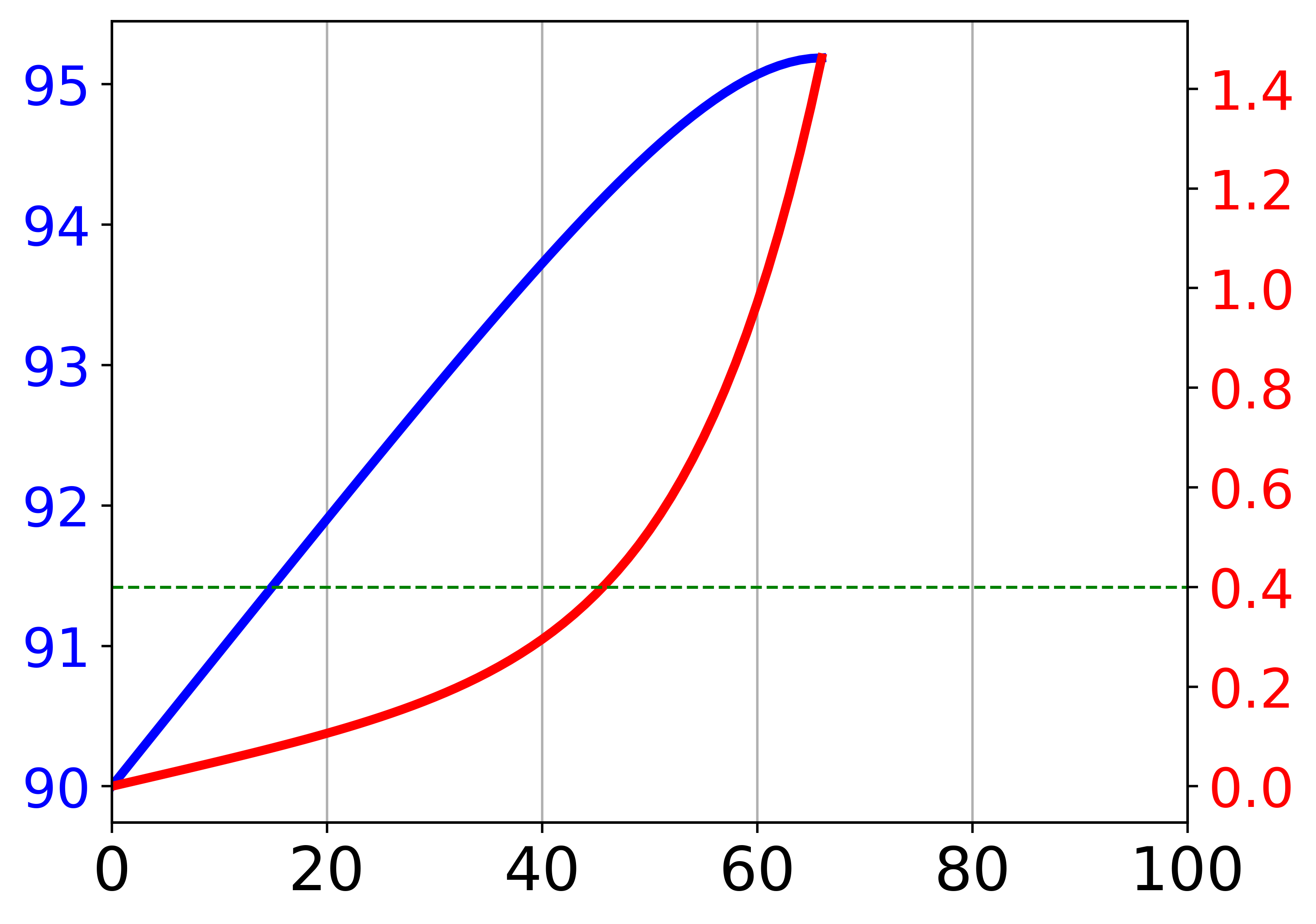}   
  \caption{}
\end{subfigure}

\caption{Pareto optimal solutions for 15 different setting of input parameters. See Table \ref{InputTable}. The horizontal axis shows the possible occupancy. The blue curve and the left vertical axis illustrate the obtained total productivity, and the red curve and the right vertical axis illustrate the expected number of infected employees per week.}
\label{ParetoSols}
\end{figure}

Furthermore, we assume two different values each for the home productivity ratio ($prod= 0.6$ and $prod= 0.9$), the vaccination rate ($0.4$ and $0.8$), the test interval ($\tau=7$ and$\tau=14$ days), and the number of contacts (low: $\kappa=5+0.10 \times occup \times n$, and high: $\kappa=5+0.20 \times occup \times n$). For $\kappa$, the low contact scenario assumes every employee in the workplace has at least 5 contacts plus additional contacts totalling 10\% of the employees present. Likewise, the high rate scenario assumes the number of contacts is at least 5 contacts plus additional contacts totalling 20\% of the employees present. Figure \ref{ParetoSols} displays the result of these 15 scenarios for different setting of input parameters, whereas Table \ref{InputTable} details the 15 input settings.

Each subfigure in Figure \ref{ParetoSols} shows Pareto optimal solutions and their objective values. The horizontal axis shows the possible occupancy values. The blue diagram and the left vertical axis illustrate the resulting total productivity, and the red diagram and the right vertical axis illustrate the expected number of infected employees per week. Note that the Pareto optimal solutions are displayed only for the occupancy values which result in such solutions, and the dominated solutions are not displayed. These results together provide useful information for a decision-maker who may consider a maximum infection risk threshold and try to find the maximum productivity possible without exceeding that threshold by changing parameters such as the vaccination rate, test interval, or contact rate. 

For example, when the 7-day incidence is 500 individuals per 100,000 population (this is almost the average number of incidences in Saxony, Germany from January first till mid February 2022, e.g., see https://www. where2test.de/saxony), the \textit{background risk} of infection per employee is $\frac{5}{1000}$ per week, and for the whole organization it is $1-(1-\frac{500}{100000})^{100} \approx 0.4$. The background risk can be interpreted as the risk of infection if the employees do not come to the office and maintain normal social contacts by, e.g., going to restaurants and shopping. When the number of incidences is intermediate, decision-makers can use the background risk as a reference point in determining a proper threshold for the risk of infection in the company. The horizontal green line in each of the subfigures displays the corresponding background risk.

For example, if a decision-maker would like to follow a workplace presence strategy where within-organization risk never exceeds the background risk, he/she can consider the occupancy corresponding to the intersection point between the background risk and the computed infection risk, i.e., the green dashed line and the red diagram. For instance, such occupancy in the first scenario in Figure \ref{ParetoSols}($a$)  ($\tau=7, prod=0.6, vaccination~rate=0.5, ~Low~contact~ rate$ and $\beta_u=0.04$) corresponds with occupancy 64\%, and it will result in 85\% productivity, while in the second scenario, Figure \ref{ParetoSols}($o$)($\tau=14, prod=0.9, vaccination~rate=0.8, ~High~contact~ rate$ and $\beta_u=0.10$), it is 46\% with more than 94\% productivity.

In a reverse usage of the Pareto optimal solutions, a decision-maker may wish to know what is the risk of infection (e.g., compared to the background risk) if he/she would like to achieve a particular level of productivity in the company. For instance, in the second scenario, \ref{ParetoSols}($b$)($\tau=7, prod=0.6, vaccination~rate=0.8, ~Low~contact~ rate$ and $\beta_u=0.04$), a level of 70\% productivity can be achieved with only half of the background risk, i.e., 0.2, resulting in only 25\% of employees present in the workplace. To compute this, we first find the intersection point of 70\% productivity with the productivity curve (blue). In the second scenario, it results in almost 25\% occupancy. Then find the infection risk on the red curve which corresponds with such occupancy, which in this case is almost 0.2.

Finally, decision makers can use several curves together to get a sense of the productivity or infection risk associated with changing key parameters, such as test interval and/or vaccination rate among employees. For instance, if the current situation in an organization is similar to the $13^{th}$ scenario (see Figure \ref{ParetoSols}($l$)), then 54\% occupancy will result in infection risk equal to the background risk with 81\% productivity. Now, if the test interval among the employees rises to $\tau=14$ days, (see 15-th scenario, Figure\ref{ParetoSols}($l$)), and the decision maker still would like the risk of infection in the organization to not exceed the background risk, he/she has to decrease occupancy to almost 46\%, which results 78\% productivity. Similar analyses can be performed to investigate the effects of other input parameters. As prevously mentioned, because of the extremely large number of possible combinations of the input parameters, it is not possible to concisely discuss all sensitivity analyses and observe changes in the objectives across the full parameter space. Instead, we suggest using the online application we provide for this purpose.

\begin{table}[]
\caption{15 different settings of input parameters. The corresponding Pareto optimal solutions are depicted in Figure \ref{ParetoSols}.}
\tiny
\label{InputTable}
\begin{center}
\begin{tabular}{|l|l|l|l|l|l|}
\hline
Figure   & $\tau$ & \textit{prod} & vaccine rate & contact rate & $\beta_u$ \\ \hline
Fig. \ref{ParetoSols}(a)  & 7   & 0.6  & 0.5    & Low     & 0.04  \\ \hline
Fig. \ref{ParetoSols}(b)  & 7   & 0.6  & 0.8    & Low     & 0.04  \\ \hline
Fig. \ref{ParetoSols}(c)  & 14  & 0.9  & 0.5    & Low     & 0.04  \\ \hline
Fig. \ref{ParetoSols}(d)  & 14  & 0.6  & 0.8    & Low     & 0.04  \\ \hline
Fig. \ref{ParetoSols}(e) & 7   & 0.9  & 0.5    & High    & 0.04  \\ \hline
Fig. \ref{ParetoSols}(f) & 14  & 0.6  & 0.5    & High    & 0.04  \\ \hline
Fig. \ref{ParetoSols}(g) & 14  & 0.9  & 0.5    & High    & 0.04  \\ \hline
Fig. \ref{ParetoSols}(h) & 7   & 0.9  & 0.5    & Low     & 0.1   \\ \hline
Fig. \ref{ParetoSols}(i) & 7   & 0.9  & 0.8    & Low     & 0.1   \\ \hline
Fig. \ref{ParetoSols}(j) & 14  & 0.6  & 0.5    & Low     & 0.1   \\ \hline
Fig. \ref{ParetoSols}(k) & 14  & 0.9  & 0.8    & Low     & 0.1   \\ \hline
Fig. \ref{ParetoSols}(l) & 7   & 0.6  & 0.8    & High    & 0.1   \\ \hline
Fig. \ref{ParetoSols}(m) & 14  & 0.9  & 0.5    & High    & 0.1   \\ \hline
Fig. \ref{ParetoSols}(n) & 14  & 0.6  & 0.8    & High    & 0.1   \\ \hline
Fig. \ref{ParetoSols}(o) & 14  & 0.9  & 0.8    & High    & 0.1   \\ \hline
\end{tabular}
\end{center}
\end{table}



\section{Conclusion and Future Directions}
In this study, we developed a model to compute the optimal trade-off solutions for minimizing the risk of infection and maximizing the productivity in organizations during the COVID-19 pandemic. First, we proposed a probabilistic approach to compute the expected number of infected employees over the time by incorporating basic influential parameters such as the local incidence level, number of contacts among the employees, the average test interval, and the vaccination rate. 
This basic probabilistic model can easily be extended to compute the number of infected individuals and probability of infection over time for groups of people differing in terms of their probabilities of infection. Though our focus here has been on COVID-19, our model and methods could easily be adapted to optimize workplace occupancy in the context of other infectious diseases.

We assumed two groups of employees with different infection probabilities (vaccinated and unvaccinated), and presented a practical approach to compute the Pareto optimal presence rate of employees to reach the maximum productivity and minimum infected risk. In addition to the incidence level, the key parameters that influence the maximum productivity are the home productivity ratio of the employees, the contact rate between the employees, the average test interval among the employees and the vaccination rate. We designed the model to be as simple as possible while still being able to cover and interpret the effects of all influential parameters. Our approach is linear in terms of time complexity, and it can be simply extended to consider the sensitivity of COVID-19 diagnostic tests. Furthermore, the model can be extended to more than two groups of employees, e.g, different age groups. A basic implementation of this model and the optimization algorithms are available online at \textit{https://test.where2test.de}. Future extensions of this study may consider the following subjects

\begin{itemize}
\item In addition to employees, visitors (i.e., clients and customers) who have direct contact with the employees can be considered for computing the probability of an infection arrival at the facility.
\item Heterogeneous groups of employees with different contact rates, physical networks, productivity ratios and test frequencies can be considered.
\item For simplicity, we defined the total productivity of the organization based on the individual productivity, that is, the sum of all employees' productivity. However, it is possible for some organizations to define their total productivity based on the type of the tasks and services in different sections of the company. To this end, the physical graph of contacts and more details on the outcomes would be required as input.
\end{itemize}

\section*{Acknowledgements}
This work was partially funded by the Center of Advanced Systems Understanding (CASUS), which is financed by Germany's Federal Ministry of Education and Research (BMBF) and by the Saxon Ministry for Science, Culture, and Tourism (SMWK) with tax funds on the basis of the budget approved by the Saxon State Parliament.
\bibliography{bibfile}

\newpage
\section*{Appendix: Evaluation of the proposed probabilistic approach}
\begin{figure}[p]
\begin{subfigure}{.5\textwidth}
  \centering
  \includegraphics[width=1 \linewidth]{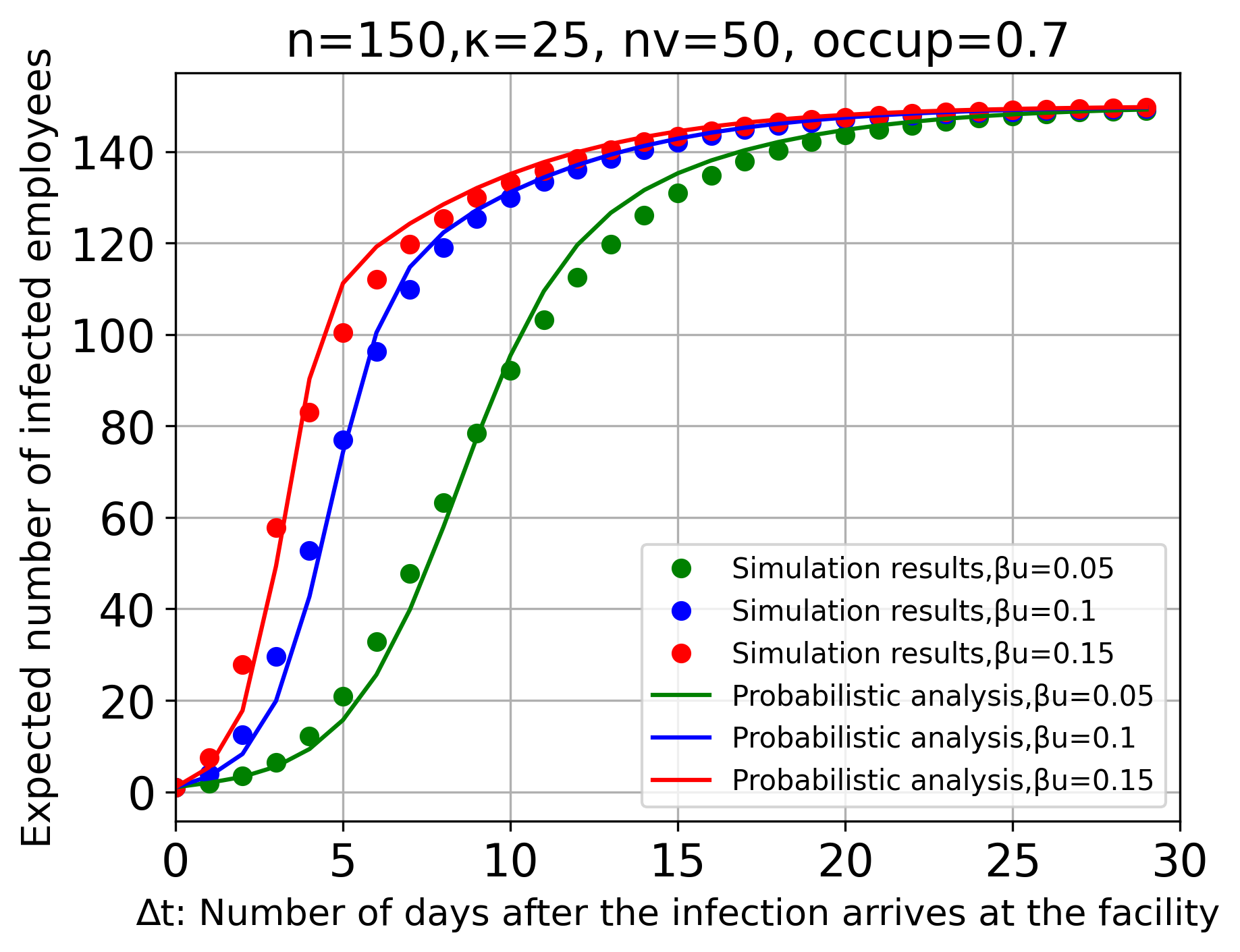}  
  \caption{}
\end{subfigure}
\begin{subfigure}{.5\textwidth}
  \centering
  \includegraphics[width=1 \linewidth]{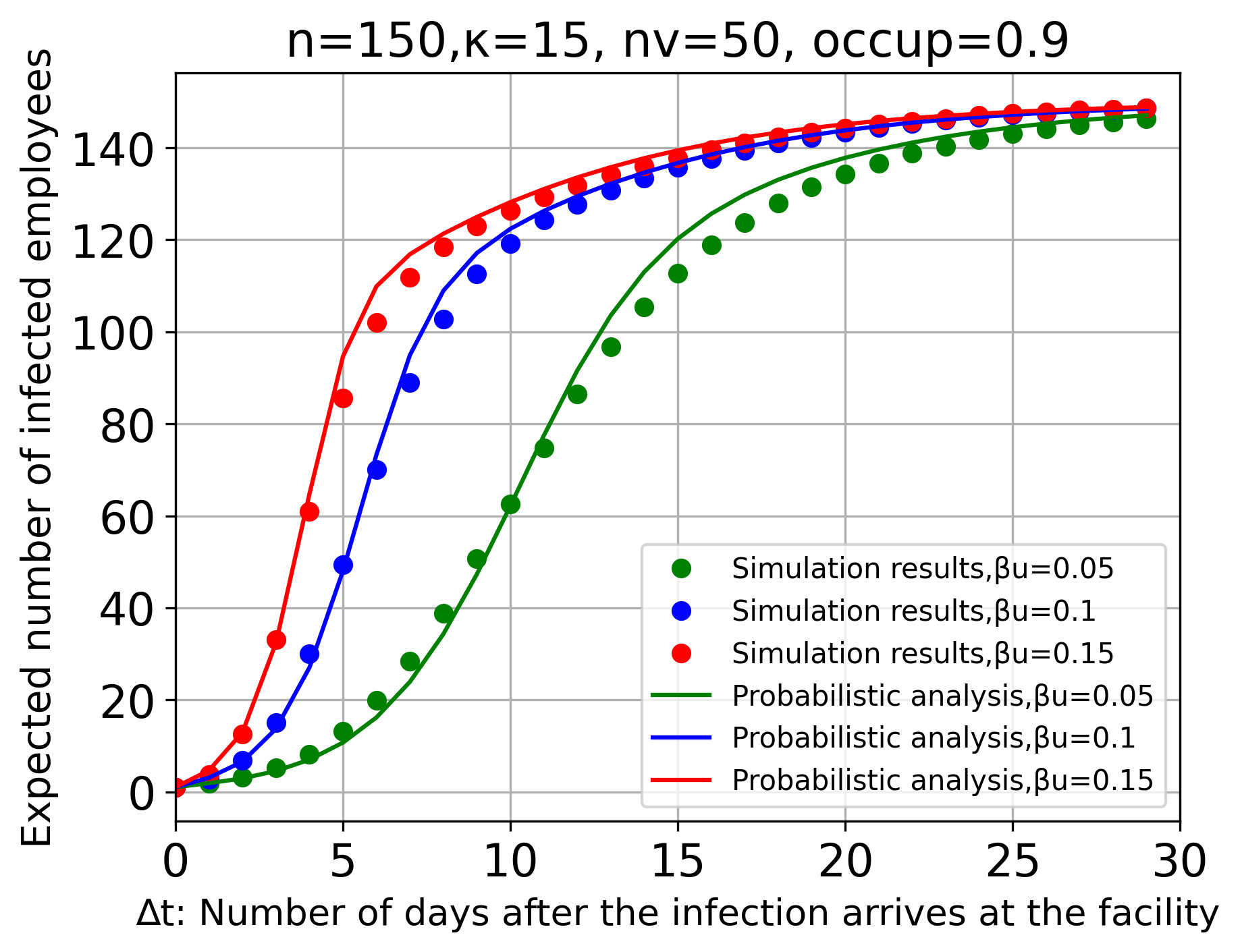}  
  \caption{}
\end{subfigure}
\begin{subfigure}{.5\textwidth}
  \centering
  \includegraphics[width=1 \linewidth]{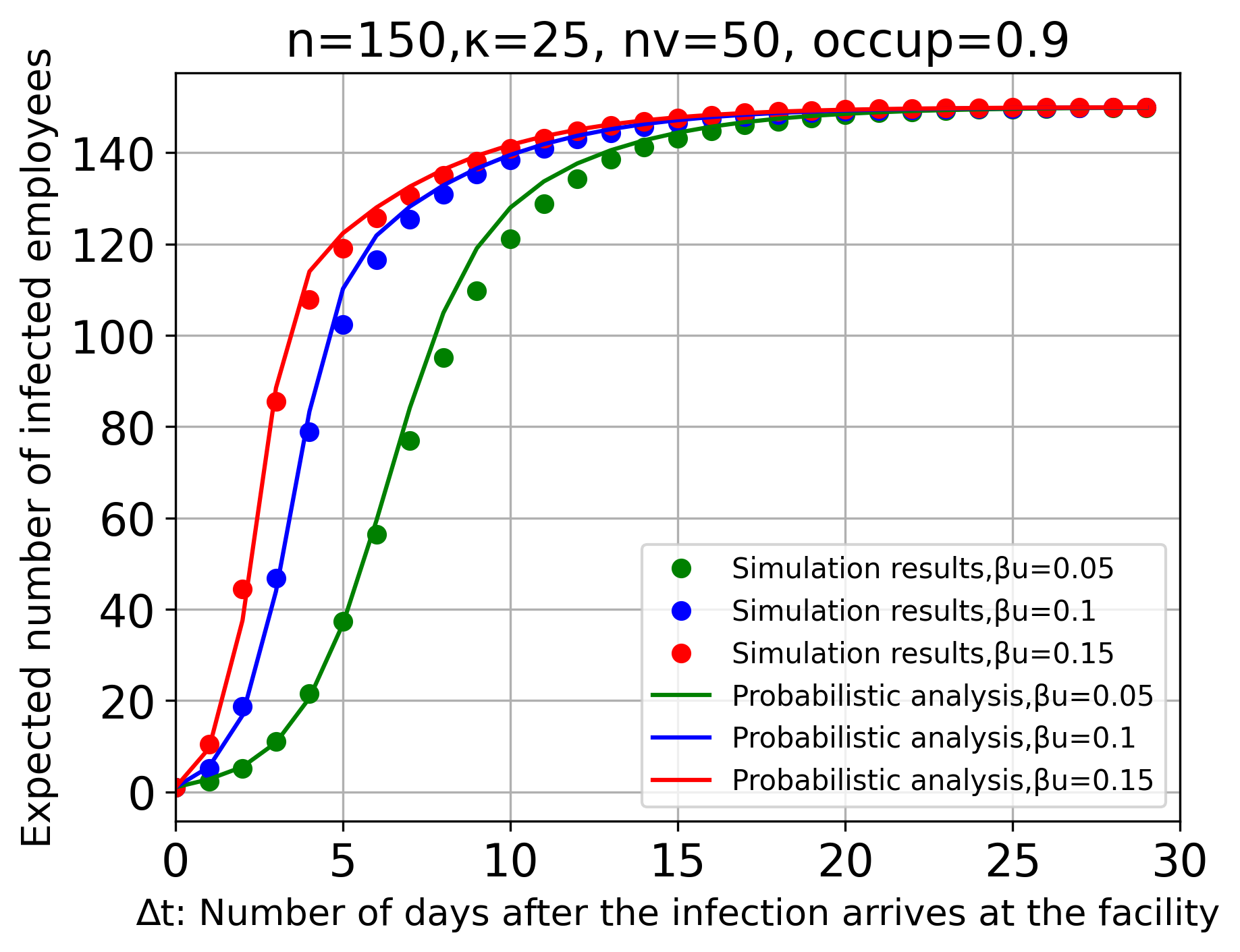}  
 \caption{}
\end{subfigure}
\begin{subfigure}{.5\textwidth}
  \centering
  \includegraphics[width=1 \linewidth]{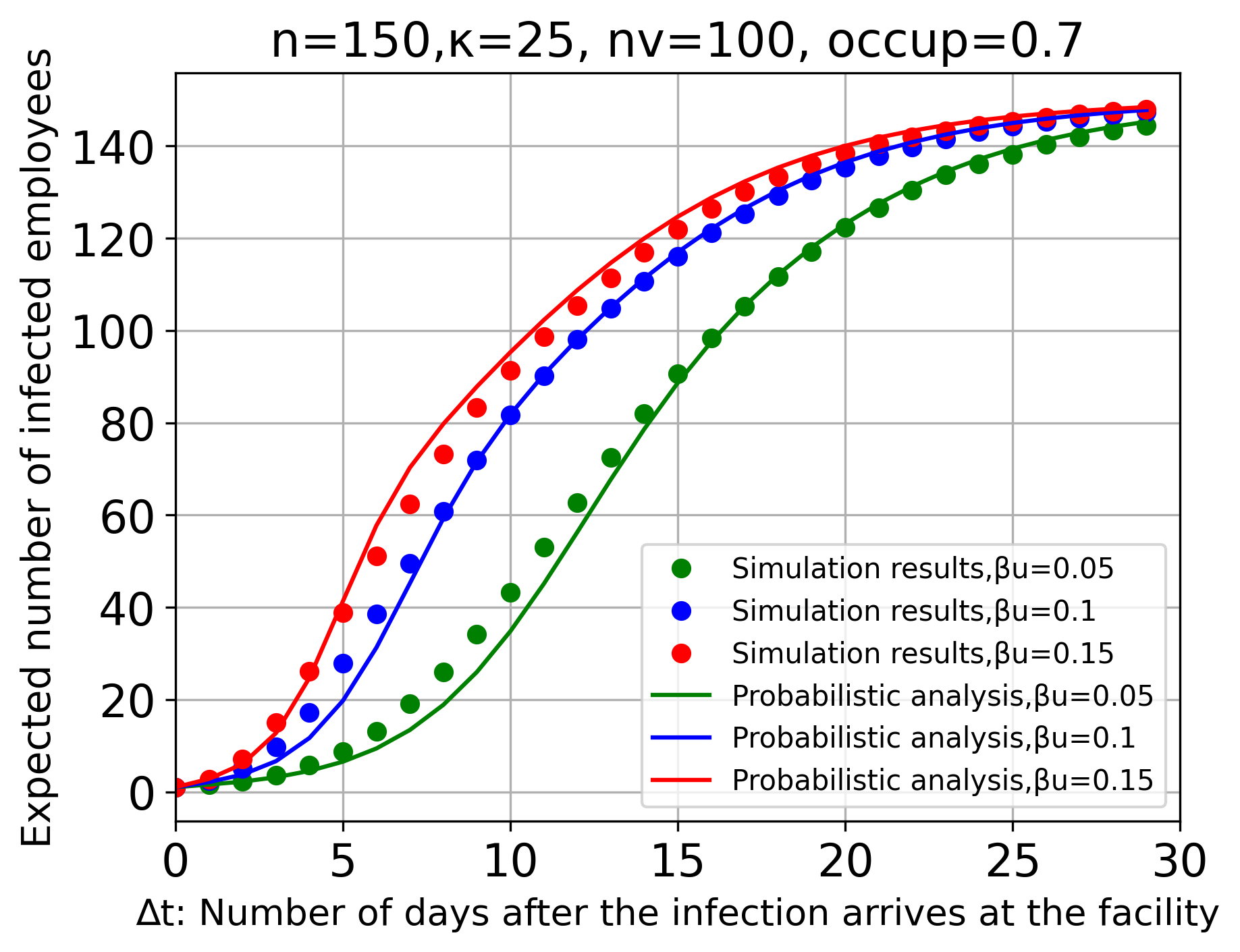}  
  \caption{}
\end{subfigure}
\begin{subfigure}{.5\textwidth}
  \centering
  \includegraphics[width=1 \linewidth]{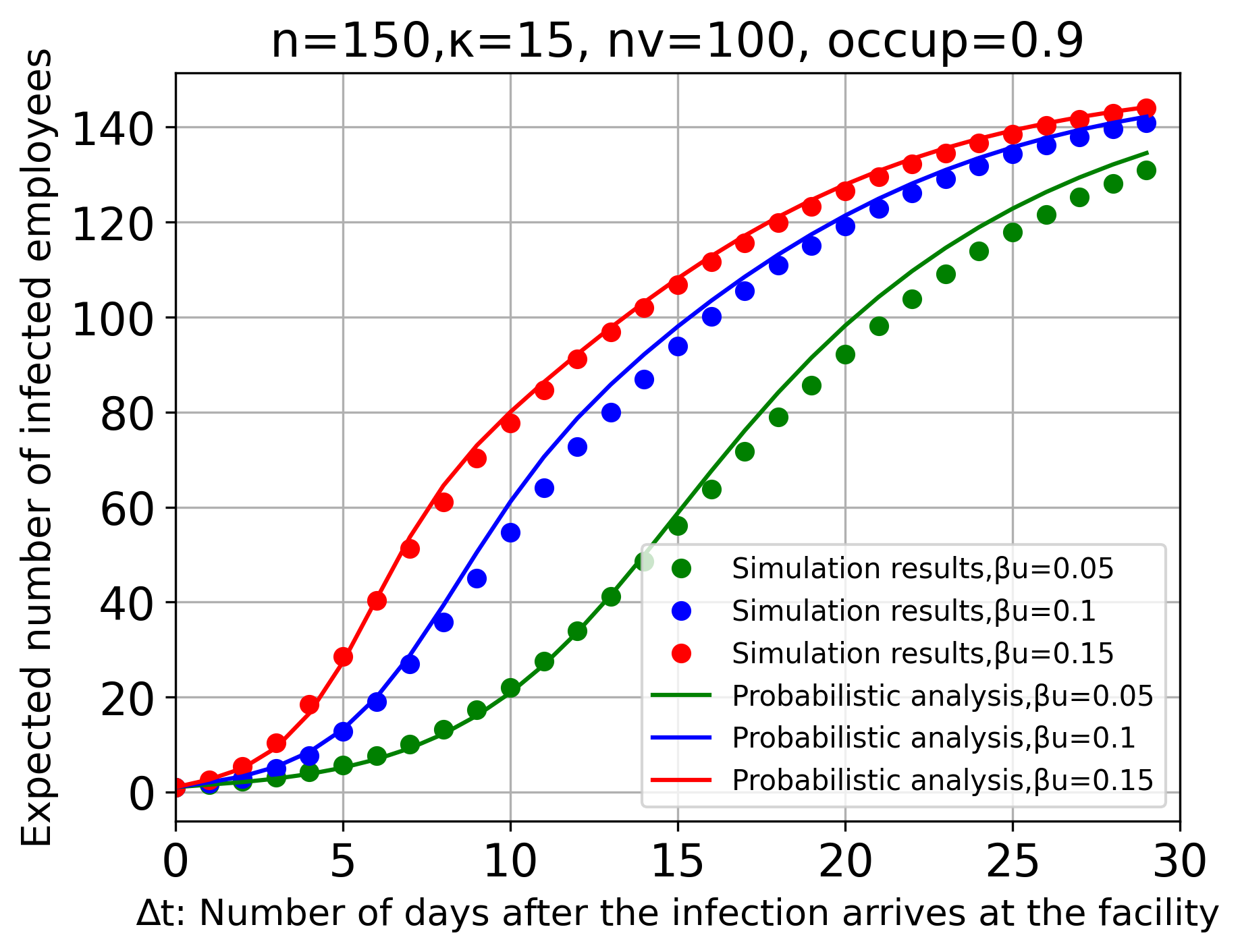}  
  \caption{}
\end{subfigure}
\begin{subfigure}{.5\textwidth}
  \centering
  \includegraphics[width=1 \linewidth]{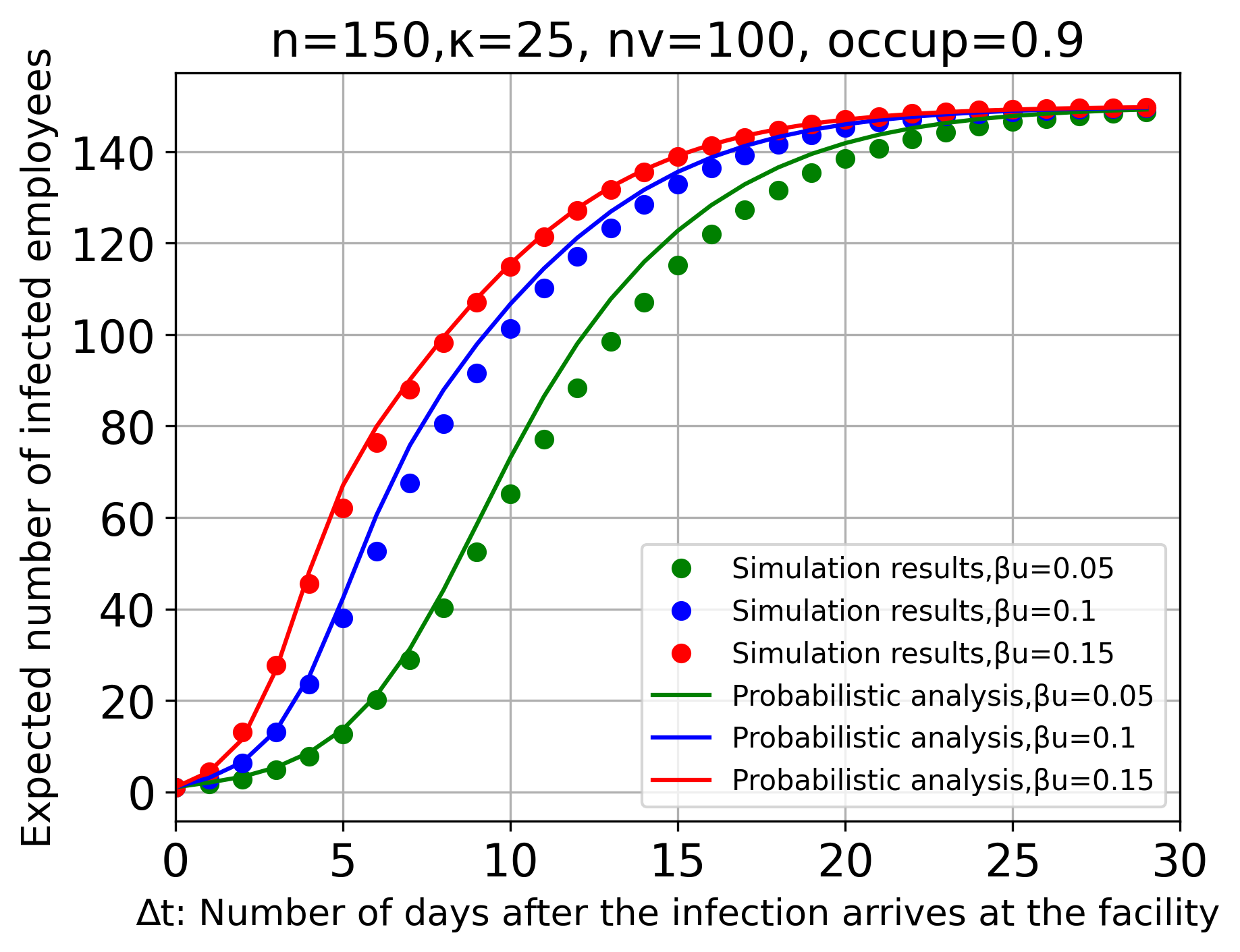}  
  \caption{}
\end{subfigure}
\caption{The simulation results and the probabilistic analysis of the expected number of infected employees for a facility with $n=150$ employees along time, $\Delta t=1,2,\dots,29$ days, after the infection arrives. Six different settings of the influential parameters are shown. The computations are based on Eq.  \ref{Vaccin_Exp_n_I}.}
\label{vac_res}
\end{figure}

However, we explained the theory behind the probabilistic approach, in this section, we implement the proposed probabilistic approach for computing the expected number of infected employees as well. To this end, we assume different settings of input parameters, and compute the expected number of infections along the time. Indeed, we suppose one infection at time $t=0$, and compute the expected infections for one month, i.e., $t < 30$ using Eq. \ref{Vaccin_Exp_n_I}. Fig. \ref{vac_res} illustrates the results for different values of $\beta_u$, $\kappa$, $occup$ and $n_v$. In these results, we assumed $n=150$ employees and $\beta_v=0.15\beta_u$. In each subfigure, three cases of transmission rates, $\beta_u=0.05$, $\beta_u=0.10$ and $\beta_u=0.15$, are shown. Also, for comparison, we show simulation results (the dots), that is, simulating the companies using a set of $n=150$ agents (employees) based on the parameter's setting. To reduce the effect of random number generators, we ran the simulations for 100 times and report the average number of infected agents. Table \ref{Sim_Comp} shows the \textit{mean absolute percentage error} for each figure and the average of them. As it can be seen, the average (the last row of the Table) percentage of difference between the results is 5.6\%, 3.8\% and 2.7\% for transmission rates $\beta_u=0.05$, $\beta_u=0.10$ and $\beta_u=0.15$, respectively. The results show the probabilistic analysis estimates the number of infected individuals with high accuracy, while it is fast and flexible approach to apply for heterogeneous group of people with different transmission rate and number of contacts.

\begin{table}[h!]
  \begin{center}
    \caption{Mean absolute percentage error between the simulation results (average of 100 iterations) and the proposed probabilistic analysis for estimating  the number of infected employees for six different settings of the influential parameters.}
    \label{Sim_Comp}
    \begin{tabular}{c c c c} \hline
      \textbf{} & \textbf{~~$\beta_u=0.05$~~~} & \textbf{$\beta_u=0.10$~~~} & \textbf{$\beta_u=0.15$}\\\hline
     	{Fig. \ref{vac_res}(a)} & {0.055} & {0.042} & {0.041}\\
	{Fig. \ref{vac_res}(b)} & {0.057} & {0.024} & {0.026}\\
	{Fig. \ref{vac_res}(c)} & {0.028} & {0.017} & {0.014}\\
	{Fig. \ref{vac_res}(d)} & {0.078} & {0.057} & {0.040}\\
	{Fig. \ref{vac_res}(e)} & {0.051} & {0.053} & {0.024}\\
	{Fig. \ref{vac_res}(f)} & {0.064} & {0.036} & {0.014}\\
      \textbf{Average} & \textbf{0.056} & \textbf{0.038} & \textbf{0.027}\\
    \end{tabular}
  \end{center}
\end{table}


\end{document}